\documentclass[prd,10pt,aps]{revtex4-2}
\usepackage{graphicx,epstopdf}
\usepackage[caption=false]{subfig}
\usepackage{appendix}
\usepackage[T1]{fontenc}
\usepackage{lmodern}
\usepackage[dvipsnames,x11names]{xcolor}
\usepackage[colorlinks=true,linkcolor=Magenta!80!black,citecolor=Magenta!70!black,urlcolor=Magenta!80!black]{hyperref}
\usepackage[sort&compress]{natbib}
\usepackage{morefloats}
\usepackage[pdf]{pstricks}
\usepackage{amsmath}
\usepackage{amssymb}
\usepackage{amsfonts}
\usepackage{rotating}
\usepackage{cancel}
\usepackage{mathtools}
\usepackage{bbm}
\usepackage{dsfont}
\usepackage{bbold}
\usepackage{multirow}
\usepackage{ulem}
\usepackage{orcidlink}
\usepackage{colortbl}
\usepackage{booktabs}
\usepackage{bm}
\usepackage{array}
\usepackage{slashed}
\usepackage{orcidlink}

\newcolumntype{d}[1]{D{.}{.}{#1}}

\newcommand{\hT}{\tilde{H}_T}
\newcommand{\eT}{\tilde{E}_T}
\newcommand{\GeV}{\ \mathrm{GeV}}
\newcommand{\GeVsq}{\ \mathrm{GeV}^2}

\begin{document}
\title{Flavour decomposition of the chiral-odd form factors of the nucleon}

\author{Ula\c{s} \"Ozdem \orcidlink{0000-0002-1907-2894}}
\email[]{ulasozdem@aydin.edu.tr}
\affiliation{Health Services Vocational School of Higher Education,
             Istanbul Aydin University, Sefakoy-Kucukcekmece,
             34295 Istanbul, T\"urkiye}


\begin{abstract}
We compute the chiral-odd form factors $H_T$, $E_T$ and $\hT$ of the nucleon within light-cone QCD sum rules, separately for the $u$ and $d$ quarks, over $1\le Q^2\le10$~GeV$^2$, and organise their values at $Q^2=0$ into the tensor monopole, dipole and quadrupole moments $g_T$, $\kappa_T$ and $Q_T$. The flavour components of the quadrupole moment, which has no chiral-even counterpart at leading twist, have not previously been obtained in this framework, nor has the isoscalar channel, for which sum rules were written down but never evaluated. The form factors are read from a basis of Lorentz structures that becomes independent only after a canonical ordering of the Dirac strings; of the eight structures that survive, three suffice to give the three form factors separately, and the structure from which the tensor charge is usually extracted is not one of the eight. Two statements follow from the analytic expressions. The $d$-quark contributions to $E_T$ and $\hT$ are equal and opposite in the chiral limit, the breaking being proportional to the quark mass and to the twist-six amplitude $\mathcal{V}_6$, so that the $d$-quark sector carries a single independent function; and, at the level of the least suppressed momentum-fraction integral, the isoscalar tensor charge receives no leading-twist contribution, the twist-three terms of the two flavour channels cancelling against one another. Neither statement is visible without resolving the flavours. At $\mu^2=1$~GeV$^2$ we obtain, for the two distribution-amplitude sets, the isovector moments $g_T^{u-d}=1.27(7)$ and $1.08(6)$, $\kappa_T^{u-d}=1.35(62)$ and $1.58(59)$, $Q_T^{u-d}=-4.40(83)$ and $-4.20(73)$, and the isoscalar moments $g_T^{u+d}=0.39(7)$ and $0.38(6)$, $\kappa_T^{u+d}=4.47(62)$ and $4.32(59)$, $Q_T^{u+d}=2.41(53)$ and $1.56(30)$. The mean-field relation $2\hT^{u-d}=-E_T^{u-d}$, tested here from a framework that makes no large-$N_c$ assumption, holds in sign and in order of magnitude, the ratio coming out $0.64(16)$ and $0.67(15)$ against the predicted unity. Carried to the impact-parameter plane, the moments displace the two flavour distributions in opposite transverse directions by nearly equal amounts, $\langle b_y\rangle=0.37$ and $-0.37$~fm, and elongate both across the polarisation axis, the $d$ quark some five to six times more strongly than the $u$ quark.
\end{abstract}

\maketitle

\section{Introduction}
\label{sec:intro}

The internal structure of the nucleon is encoded in the matrix elements of local
QCD operators. Those of the vector and axial-vector currents have been mapped in
detail; the third leading-twist structure, the one probed by the tensor current,
remains the least constrained. It measures how quarks are polarised transversely
to the nucleon spin, and both the experimental and the theoretical
determinations of it are still sparse. The present work computes the tensor form
factors of the nucleon within QCD light-cone sum rules, resolved into $u$- and
$d$-quark contributions, and organises them into the three multipole moments
that carry the physical content: the tensor charge, the anomalous tensor
magnetic moment and the tensor quadrupole moment. Resolving the flavours is not
a refinement here but the substance of the calculation, since each of the
structural results reported below lives in one flavour channel or in the
interplay of the two.

The tensor current $\bar{q}\,i\sigma_{\mu\nu}q$ is odd under charge conjugation,
so its forward matrix element, the tensor charge $\delta q$, measures the
difference between the transverse polarisation of quarks and that of
antiquarks: sea quark--antiquark pairs cancel in it to the extent that quarks
and antiquarks are equally polarised, and $\delta q$ is in this sense a valence
quantity rather than a measure of the full quark content of the
nucleon~\cite{Jaffe:1991kp,Jaffe:1991ra}. The axial-vector
current behaves in the opposite way: it is C-even, sea-quark helicities survive,
and the singlet axial charge is correspondingly difficult to interpret. The
tensor charge is the first moment of the transversity distribution $h_1$, the
last of the three leading-twist collinear distributions to be
mapped~\cite{Ralston:1979ys,Artru:1989zv,Jaffe:1991kp,Cortes:1991ja,Barone:2001sp,Barone:2010zz}: being
chiral-odd, $h_1$ decouples from inclusive deep inelastic scattering and must be
paired with a second chiral-odd object, which has restricted its extraction to
semi-inclusive processes and to dihadron production, and it is constrained from
above by the Soffer inequality~\cite{Soffer:1994ww}.
Interest in the tensor charge is not confined to hadron structure. The
flavour-diagonal tensor charges are precisely the matrix elements that convert a
quark electric dipole moment into a neutron
one~\cite{Pospelov:2005pr,Bhattacharya:2015esa,Pitschmann:2014jxa}, so a bound on
the neutron electric dipole moment translates into a bound on CP-violating
physics beyond the Standard Model only to the accuracy with which $g_T^{q}$ is
known~\cite{Liu:2017olr,Courtoy:2015haa}. The same matrix elements control the
sensitivity of precision $\beta$-decay measurements to BSM tensor
interactions~\cite{Bhattacharya:2011qm}.

Experimentally, transversity and the tensor charge have been extracted from
Collins asymmetries in semi-inclusive deep inelastic scattering measured by
HERMES~\cite{Airapetian:2010ds} and COMPASS~\cite{Adolph:2012sn}, combined with
$e^{+}e^{-}$ data from Belle~\cite{Vossen:2011fk}, and from dihadron
production~\cite{Bacchetta:2012ty,Radici:2015mwa}; global analyses of increasing
scope have followed~\cite{Anselmino:2013vqa,Kang:2015msa,Lin:2017stx,Cocuzza:2023oam}.
The precision remains well below that of the helicity sector,
and improving it is among the goals of the Electron-Ion
Collider~\cite{Accardi:2012qut,AbdulKhalek:2021gbh}.
On the theory side the tensor charge has been computed in lattice
QCD~\cite{Gockeler:2005cj,Bhattacharya:2015wna,Gupta:2018lvp,Alexandrou:2019brg,Alexandrou:2022dtc}, in the chiral quark-soliton
model~\cite{Ledwig:2010tu,Ledwig:2010zq,Ghim:2025gqo,Ghim:2026qnp}, in the bag
model~\cite{Tezgin:2024tfh}, in light-front and covariant quark
models~\cite{Schmidt:1997vm,Pasquini:2005dk,Cloet:2007em,Maji:2017bcz}, in
Dyson--Schwinger approaches~\cite{Pitschmann:2014jxa,Xu:2015kta}, and in QCD sum
rules~\cite{He:1994gz,Erkol:2011iw,Aliev:2011ku}. The tensor form factors of
other hadronic systems have also been
studied~\cite{Ozdem:2020vpt,Ozdem:2021zbn,kucukarslan:2016xhx,Asmaee:2025elo,Asmaee:2026wrk}.

At leading twist the matrix element of the tensor current between spin-$1/2$
states is parametrised by three form factors, $H_T$, $E_T$ and $\hT$, the first
Mellin moments of the chiral-odd generalised parton
distributions~\cite{Diehl:2001pm,Diehl:2005jf}. As such they are the natural
variables in momentum space, but for the spatial interpretation the physical
content is carried by particular combinations of them. Expanding the matrix element in irreducible tensors built
from the momentum transfer separates a monopole, a dipole and a quadrupole term,
an enumeration carried out systematically for the chiral-odd operator in
Ref.~\cite{Kim:2024ibz}, where the quadrupole structure was identified as a term
that earlier analyses had omitted. The monopole is the tensor charge. The dipole
moment $\kappa_T$ measures the shift of the average transverse position of
quarks polarised in the transverse direction, in an unpolarised
target~\cite{Burkardt:2005hp}. That distortion, combined with final-state
interactions, has been argued to generate the Boer--Mulders
function~\cite{Burkardt:2005hp}, so that the sign of $\kappa_T$ carries a
phenomenological consequence; the chiral-odd distributions moreover allow the
quark angular momentum to be decomposed with respect to quarks of definite
transversity, giving access to the correlation between quark spin and orbital
angular momentum in an unpolarised nucleon~\cite{Burkardt:2005hp}. The
quadrupole moment $Q_T$ characterises the deformation of the corresponding
distribution when the nucleon is transversely polarised as well, and is peculiar
to the chiral-odd sector: at leading twist the chiral-odd operator carries
$L_3=\pm1$, and combined with the rank-two tensor the quadrupole is the highest
structure that can be formed, whereas the chiral-even operator has $L_3=0$ and
cannot reach it at all~\cite{Diehl:2005jf}. The chiral-odd generalised parton
distributions themselves have been studied on the lattice~\cite{Gockeler:2006zu}
and in a range of models~\cite{Pasquini:2005dk, Maji:2017bcz, Kaur:2023lun, Tezgin:2024tfh}.
Their measurement meets the same obstacle as that of $h_1$, a second chiral-odd
object being required, and exclusive processes in which this role is played by
the distribution amplitude of a transversely polarised $\rho$ meson have been
proposed, in the diffractive electroproduction of two vector
mesons~\cite{Ivanov:2002jj} and in the photoproduction of a $\gamma\rho$ pair at
large invariant mass~\cite{Boussarie:2016qop}.

Within light-cone QCD sum rules the coverage is uneven. The isovector form
factors were obtained in Ref.~\cite{Erkol:2011iw}; Ref.~\cite{Aliev:2011ku}
derived the sum rules for both isovector and isoscalar currents using the
general interpolating field, but quoted no numbers, having found that the form
factors change sign with the mixing parameter. Neither study formed the
multipole combinations. The isovector value of $2\hT(0)$ is implied by the
$\hT^{u-d}(0)$ of Ref.~\cite{Erkol:2011iw}, and we construct it for the
comparison of Sec.~\ref{sec:results}, but the quadrupole moment is not discussed
there, and neither its flavour components nor its isoscalar combination have
been obtained in this framework. 
The present work differs from these studies in four respects. The form factors
are computed separately for the $u$ and $d$ quarks, and both isospin
combinations are evaluated, the isoscalar one included. The flavour resolution
requires no new technique, each flavour component of the tensor current being
inserted into the correlator separately; what makes it informative is that, with
the Ioffe current written with the two $u$ quarks adjacent, the $u$- and
$d$-quark contractions draw in the chiral limit on disjoint sets of distribution
amplitudes, and the two structural results of Secs.~\ref{sec:isoscalar}
and~\ref{sec:identity} follow from that separation. The three multipole moments,
the quadrupole included, are formed in every flavour channel. The form factors
are read from a basis of Lorentz structures that is independent only after a
canonical ordering of the Dirac strings; before ordering the structures are
linearly related, so the coefficient assigned to any one of them depends on how
the expression happens to have been written. And the results are used to test a
relation derived in the mean-field picture, from a framework that makes no
assumption about $N_c$. The moments are finally carried to the impact-parameter
plane, where the three of them fix the displacement and the deformation of the
transverse quark distribution.

The paper is organised as follows. Section~\ref{sec:formalism} sets up the
multipole decomposition, fixes the independent basis and derives the sum rules.
Section~\ref{sec:numerics} describes the input, the working regions of the
auxiliary parameters and the extraction procedure. Section~\ref{sec:results}
presents the results and discusses them. Section~\ref{sec:summary} summarises.
The QCD-side expressions are collected in the Appendix.

\section{Formalism}
\label{sec:formalism}

\subsection{Form factors}
\label{sec:ff}

The matrix element of the tensor current between nucleon states is parametrised
by three form factors,
\begin{align}
\langle N(p')|\bar q\,i\sigma_{\mu\nu}q|N(p)\rangle
&=\bar u(p')\Big[\,i\sigma_{\mu\nu}\,H_T^{q}(t)
 +\frac{\gamma_\mu q_\nu-\gamma_\nu q_\mu}{2m_N}\,E_T^{q}(t)+\frac{\tilde P_\mu q_\nu-\tilde P_\nu q_\mu}{2m_N^{2}}\,\hT^{q}(t)
 \Big]u(p),
\label{eq:param}
\end{align}
with $\tilde P=p'+p$, $q=p-p'$ and $t=q^2=-Q^2$; $q$ is the momentum flowing
into the tensor current, so that $p'=p-q$. A fourth structure is allowed
by Lorentz covariance, but the corresponding form factor $\eT$ vanishes
by T-invariance~\cite{Aliev:2011ku,Diehl:2001pm}. The three are the first Mellin
moments of the chiral-odd generalised parton distributions.

\subsection{Multipole structure}
\label{sec:multipole}

As first Mellin moments of the chiral-odd generalised parton distributions the
three form factors are the natural variables in momentum space, but they do not
coincide one by one with the multipoles of the transverse spatial distribution:
$\hT$ enters both the dipole and the quadrupole, and $E_T$ enters the dipole only
in combination with it. The multipole content is exposed by expanding the matrix
element in irreducible tensors built from the momentum transfer, which separates
a monopole, a dipole and a quadrupole term. Expanding in the transverse plane at fixed light-front time,
which is where the density interpretation is available, gives~\cite{Kim:2024ibz}
\begin{equation}
\begin{aligned}
\text{monopole}&:\quad g_T^{q}=H_T^{q}(0),\\
\text{dipole}&:\quad \kappa_T^{q}=E_T^{q}(0)+2\hT^{q}(0),\\
\text{quadrupole}&:\quad Q_T^{q}=2\hT^{q}(0) .
\end{aligned}
\label{eq:2d}
\end{equation}
The monopole $g_T$ is the tensor charge, the net number of transversely
polarised quarks in a transversely polarised nucleon. The dipole $\kappa_T$
measures the shift of the average transverse position of quarks polarised along
a fixed transverse direction, in an unpolarised target~\cite{Burkardt:2005hp}
--- the one multipole that survives when the nucleon polarisation is averaged
over; it plays for the tensor current the role that the anomalous magnetic
moment plays for the vector one. The quadrupole $Q_T$ characterises the
deformation of the transverse distribution when the nucleon is polarised as
well~\cite{Diehl:2005jf}, a configuration developed in Sec.~\ref{sec:shapes}. A
second, three-dimensional expansion in the Breit frame is also in
use~\cite{Ghim:2026qnp}; the two sets do not coincide, differing by terms of
order $g_T$, and every result below refers to Eq.~\eqref{eq:2d}.

The expansion stops at the quadrupole. At leading twist the chiral-odd operator
carries $L_3=\pm1$; combined with the rank-two tensor built from the momentum
transfer, the quadrupole is the highest structure that can be formed. The
chiral-even operator has $L_3=0$ and cannot reach the quadrupole at all, so this
term is peculiar to the chiral-odd sector.

\subsection{Light-cone QCD sum rules}
\label{sec:lcsr}

The starting point is
\begin{equation}
\Pi_{\mu\nu}(p,q)=i\!\int\!d^{4}x\,e^{iqx}
 \langle0|\mathcal{T}\{J_N(0)\,\bar q\,\sigma_{\mu\nu}q(x)\}|N(p)\rangle .
\label{eq:corr}
\end{equation}
The interpolating field is not unique: the general local three-quark operator
with the quantum numbers of the nucleon carries a free mixing parameter, and the
resulting form factors depend on it. We fix it to the Ioffe value throughout,
so that
\begin{equation}
J_N=2\epsilon^{abc}\big[(u^{aT}Cd^{b})\gamma_5u^{c}
 -(u^{aT}C\gamma_5d^{b})u^{c}\big]
 =\epsilon^{abc}\big(u^{aT}C\gamma_\lambda u^{b}\big)\gamma_5\gamma^{\lambda}d^{c} .
\label{eq:ioffe}
\end{equation}
Three reasons. First, this is the choice for which the coupling of the current
to the nucleon ground state is largest relative to the continuum in the
two-point function, the criterion originally proposed in
Ref.~\cite{Ioffe:1981kw}. Second, it is the current adopted in the analysis from
which the distribution amplitudes used here are taken~\cite{Braun:2006hz}, so
that the current and its nonperturbative input are matched. Third, and specific
to the present calculation, only at this value can the current be written with
the like-flavour pair adjacent, as in the second form of Eq.~\eqref{eq:ioffe},
and it is in that form that the selection rule derived below becomes manifest.

The two earlier light-cone QCD sum rule studies of the tensor form factors work at
other values of the mixing parameter. Writing it as $\beta=\tan\theta$, the
Ioffe current corresponds to $\cos\theta\simeq0.71$, whereas
Ref.~\cite{Erkol:2011iw} reports stability near $\cos\theta\simeq0$ and
Ref.~\cite{Aliev:2011ku} works in the window $-0.5\le\cos\theta\le0.3$.

In the second writing of Eq.~\eqref{eq:ioffe} the two $u$ Dirac indices sit on
the same object and the distribution amplitude carries them together, so the
chain closes and each contraction is a Dirac trace multiplying an open string.
The tensor current can be contracted with either of the two $u$ quarks of the
diquark $u^{aT}C\gamma_\lambda u^{b}$, which gives the contractions labelled (A)
and (B) below, or with the $d$ quark, which gives (C); in (B) the contracted field
is the transposed one, which is why the conjugated propagator $\widehat S_u$
appears there. Using $C\Gamma^{T}C^{-1}=\eta_\Gamma\Gamma$, $C^{2}=-1$ and
$C\sigma_{\mu\nu}^{T}S^{T}C=\sigma_{\mu\nu}\widehat S$ with
$\widehat S_q=CS_q^{T}C^{-1}$, every factor of $C$ cancels and
\begin{align}
\text{(A)}&=\eta_X\,
 \mathrm{Tr}\big[\widetilde\Gamma_X\gamma_\lambda S_u\sigma_{\mu\nu}\big]\,
 \big[\gamma_5\gamma^{\lambda}\Gamma_3^{X}\big],\\
\text{(B)}&=\,
 \mathrm{Tr}\big[\sigma_{\mu\nu}\widehat S_u\gamma_\lambda\widetilde\Gamma_X\big]\,
 \big[\gamma_5\gamma^{\lambda}\Gamma_3^{X}\big],\\
\text{(C)}&=\eta_X\,
 \mathrm{Tr}\big[\gamma_\lambda\widetilde\Gamma_X\big]\,
 \big[\gamma_5\gamma^{\lambda}S_d\sigma_{\mu\nu}\Gamma_3^{X}\big],
\end{align}
where $\widetilde\Gamma_X$ is the diquark Dirac structure of the amplitude $X$,
$\Gamma_3^{X}$ the structure carrying the third quark index, and $\eta_X=+1$ for
the $S$, $P$ and $A$ amplitudes and $-1$ for $V$ and $T$. In $\widehat S_q$ the
$\slashed x$ term of the propagator flips sign while the mass term does not, so
the two must be kept as separate objects once $m_q\neq0$.

The $d$-quark contraction (C) contains $\mathrm{Tr}[\gamma_\lambda\widetilde\Gamma_X]$, four
times the vector component of $\widetilde\Gamma_X$, which vanishes unless
$\widetilde\Gamma_X$ is $\slashed p$, $\gamma_\kappa$ or $\slashed x$. Only the
six $V$ amplitudes therefore contribute to the $d$-quark channel; $S$, $P$, $A$
and $T$ drop out. This selection rule is used in Secs.~\ref{sec:isoscalar}
and~\ref{sec:identity}.

The light-quark propagator is kept with its mass term,
\begin{equation}
S_q(x)=\frac{i\slashed x}{2\pi^{2}x^{4}}-\frac{m_q}{4\pi^{2}x^{2}}+\dots ,
\label{eq:prop}
\end{equation}
the condensate pieces being removed by the Borel transformation and the gluon
pieces feeding four- and five-particle amplitudes, which are
small~\cite{Diehl:1998kh,Braun:2006hz} and are neglected. The nonperturbative
input is the three-quark matrix element, whose decomposition into
distribution amplitudes of definite twist we take unchanged from
Refs.~\cite{Braun:2000kw,Braun:2006hz}; we follow their notation throughout.

On the hadronic side the nucleon contribution is isolated by inserting a
complete set of states and keeping the ground-state pole,
\begin{equation}
\Pi_{\mu\nu}=\lambda_N\,
 \frac{\slashed p'+m_N}{m_N^{2}-p'^{2}}\;\Gamma_{\mu\nu}(q)\,u(p)
 \;+\;\text{continuum},\qquad p'=p-q ,
\label{eq:hadronic}
\end{equation}
with $\langle0|J_N|N(p')\rangle=\lambda_N u(p')$ and $\Gamma_{\mu\nu}$ the
decomposition of Eq.~\eqref{eq:param}.

The two sides are now matched structure by structure, and this is where a choice
has to be made that is usually left implicit. The Dirac strings generated by the
operator product are not independent as they stand; the Clifford algebra relates
them, for instance
\begin{equation}
\slashed q\,\sigma_{\mu\nu}
 =\sigma_{\mu\nu}\slashed q+2i\big(q_\mu\gamma_\nu-q_\nu\gamma_\mu\big) ,
\label{eq:cliffordexample}
\end{equation}
so that $\slashed q\,\sigma_{\mu\nu}$, $\sigma_{\mu\nu}\slashed q$ and
$q_\mu\gamma_\nu-q_\nu\gamma_\mu$ cannot be treated as three separate objects.
The coefficient one assigns to any one of them therefore depends on how the
expression happens to have been written, not only on its content, and the same
ambiguity affects $\sigma_{\mu\nu}$, $q_\mu\gamma_\nu$, $\gamma_\mu q_\nu$ and
$g_{\mu\nu}$. A projection performed before the strings are brought to a common
arrangement is thus not well defined, whichever side of the sum rule it is
applied to.

We remove the ambiguity by reducing every string to the canonical arrangement
\begin{equation}
\gamma_\mu\,\gamma_\nu\,\slashed q\,\slashed p
\label{eq:canon}
\end{equation}
through repeated use of
\begin{equation}
\ldots y\,z\ldots \;\longrightarrow\;
2\,(y\!\cdot\!z)\,\ldots\mathbb{1}\ldots\;-\;\ldots z\,y\ldots ,
\qquad y\neq\gamma_5 ,
\label{eq:order}
\end{equation}
applied so as to move $\slashed q$, then $\gamma_\nu$, then $\gamma_\mu$
successively to the left, each substitution generating a scalar product and a
reordered string, and eliminating $\slashed p$ through $\slashed p\,u(p)=m_N
u(p)$. What remains is a set of eight mutually irreducible antisymmetric
structures with scalar coefficients built from $q_\mu$, $q_\nu$, $p_\mu$,
$p_\nu$ and $g_{\mu\nu}$. Applying the same reduction to the right-hand side of
Eq.~\eqref{eq:hadronic} gives
\begin{align}
\Pi_{\mu\nu}=\frac{\lambda_N}{m_N^{2}-p'^{2}}\Big\{\;
 &H_T\Big[2\big(q_\mu\gamma_\nu-q_\nu\gamma_\mu\big)
          -2\big(p_\mu\gamma_\nu-p_\nu\gamma_\mu\big)+2m_N\,i\sigma_{\mu\nu}-i\sigma_{\mu\nu}\slashed q\Big]\nonumber\\[4pt]
 +\;&\frac{E_T}{2m_N}\Big[\big(q_\nu\gamma_\mu-q_\mu\gamma_\nu\big)\slashed q
            -2\big(q_\mu p_\nu-q_\nu p_\mu\big)\Big]\nonumber\\[4pt]
 +\;&\hT\Big[-\frac{2}{m_N}\big(q_\mu p_\nu-q_\nu p_\mu\big)
            +\frac{1}{m_N^{2}}\big(q_\mu p_\nu-q_\nu p_\mu\big)\slashed q\Big]
 \Big\}\,u(p) .
 \label{eq:hadreduced}
\end{align}
Of the eight structures, $(p_\mu\gamma_\nu-p_\nu\gamma_\mu)\slashed q$ does not
appear at all and $q_\mu p_\nu-q_\nu p_\mu$ mixes, receiving
$-(E_T+2\hT)/m_N=-\kappa_T/m_N$ and so giving access to the dipole moment rather
than to either form factor separately. The other six are clean, but four of them
return $H_T$; the three that give one form factor each are
\begin{equation}
\begin{aligned}
q_\mu\gamma_\nu-q_\nu\gamma_\mu \;&:\;\; 2H_T ,\\
\big(q_\nu\gamma_\mu-q_\mu\gamma_\nu\big)\slashed q \;&:\;\; \frac{E_T}{2m_N} ,\\
\big(q_\mu p_\nu-q_\nu p_\mu\big)\slashed q \;&:\;\; \frac{\hT}{m_N^{2}} ,
\end{aligned}
\label{eq:projection}
\end{equation}
and these are the structures we project onto. Of the four that return $H_T$
alone we read it from $q_\mu\gamma_\nu-q_\nu\gamma_\mu$, which carries the
lowest Dirac structure of the four; the remaining three are available as a
consistency check.

Earlier light-cone QCD sum rule work extracts $H_T$ from
$\slashed q\,\sigma_{\mu\nu}$~\cite{Erkol:2011iw,Aliev:2011ku}, which by
Eq.~\eqref{eq:cliffordexample} is not one of the eight. We do not claim this
leads to an incorrect result --- both pieces on the right of
Eq.~\eqref{eq:cliffordexample} happen to carry $H_T$ alone --- but the
assignment of a QCD-side term to that structure is not unique before ordering,
and the basis used here is therefore different from theirs, independently of the
interpolating current or the fit range. This is one of the differences to bear
in mind when the results are compared in Sec.~\ref{sec:results}.

Equating the two representations, Borel transforming in $p'^{2}$ and
subtracting the continuum with the standard substitution
rules~\cite{Braun:2006hz} then gives
\begin{align}
H_T^{q}(Q^{2})&=\frac{1}{2\lambda_N}\,e^{m_N^{2}/M^{2}}\,
   \mathcal{B}_{M^2,s_0}\big[\Pi^{q}_{H_T}\big],\nonumber\\
E_T^{q}(Q^{2})&=\frac{2m_N}{\lambda_N}\,e^{m_N^{2}/M^{2}}\,
   \mathcal{B}_{M^2,s_0}\big[\Pi^{q}_{E_T}\big],
\label{eq:sumrules}\\
\hT^{q}(Q^{2})&=\frac{m_N^{2}}{\lambda_N}\,e^{m_N^{2}/M^{2}}\,
   \mathcal{B}_{M^2,s_0}\big[\Pi^{q}_{\hT}\big],\nonumber
\end{align}
where $\mathcal{B}_{M^2,s_0}$ denotes the Borel transformation together with the
continuum subtraction and $\Pi^{q}_{F}$ is the coefficient of the corresponding
structure of Eq.~\eqref{eq:projection} on the QCD side, given in
Appendix~\ref{app:rho}. The prefactors follow directly from
Eq.~\eqref{eq:projection} and are independent of $Q^{2}$; every sign in
Eq.~\eqref{eq:sumrules} follows from the convention $p'=p-q$, and with the
opposite choice the three would be reversed.

\section{Numerical analysis}
\label{sec:numerics}

\subsection{Input parameters}
\label{sec:inputs}

We use 
$m_N=0.94$ GeV, $m_u=2.16^{+0.49}_{-0.26}$ MeV and $m_d=4.67^{+0.48}_{-0.17}$ MeV~\cite{ParticleDataGroup:2024cfk}, which enter the
expressions of Appendix~\ref{app:rho} through the ratios $r_q=m_q/m_N$. The residue $\lambda_N$ is
taken from the two-point sum rule~\cite{Erkol:2011iw}.

Unlike the vector and axial-vector currents the tensor current is
renormalisation-scale dependent already at one loop. The distribution amplitudes
are normalised at $\mu^2=1\GeVsq$ and all results are quoted at that scale;
results obtained elsewhere are evolved using~\cite{Barone:2001sp}
\begin{equation}
X(\mu^2)=\left(\frac{\alpha_s(\mu^2)}{\alpha_s(\mu_i^2)}\right)^{4/27}
\left[1-\frac{337}{468\pi}\big(\alpha_s(\mu_i^2)-\alpha_s(\mu^2)\big)\right]
X(\mu_i^2),
\label{eq:evol}
\end{equation}
with
\begin{equation}
\alpha_s(\mu^2)=\frac{4\pi}{9\ln(\mu^2/\Lambda_{\rm QCD}^2)}
\left[1-\frac{64}{81}\,
\frac{\ln\ln(\mu^2/\Lambda_{\rm QCD}^2)}{\ln(\mu^2/\Lambda_{\rm QCD}^2)}\right],
\label{eq:alphas}
\end{equation}
$\Lambda_{\rm QCD}=0.25\GeV$.

We use two distribution-amplitude parameter sets, listed in
Table~\ref{tab:nucleonDA}: Set~I, in which all eight nonperturbative parameters
are estimated from QCD sum rules, and Set~II, the asymptotic set in which the
higher conformal-spin contributions are required to
vanish~\cite{Braun:2006hz,Lenz:2009ar}. The spread between them is carried in
the error budget as the distribution-amplitude systematic.

\begin{table}[htb]
\caption{Nonperturbative parameters entering the nucleon DAs~\cite{Braun:2006hz}.
The normalisation constants $f_N$, $\lambda_1$ and $\lambda_2$ are common to the
two sets.}
\label{tab:nucleonDA}
\centering
\setlength{\tabcolsep}{14pt}
\renewcommand{\arraystretch}{1.35}
\begin{tabular}{l c c}
\toprule
\textbf{Parameter} & \textbf{Set-I} & \textbf{Set-II} \\
\midrule
$f_N$         & \multicolumn{2}{c}{$(5.0 \pm 0.5) \times 10^{-3}$~GeV$^2$} \\
$\lambda_1$   & \multicolumn{2}{c}{$(-2.7 \pm 0.9) \times 10^{-2}$~GeV$^2$} \\
$\lambda_2$   & \multicolumn{2}{c}{$(5.4 \pm 1.9) \times 10^{-2}$~GeV$^2$} \\
\midrule
$A_1^u$       & $0.38 \pm 0.15$ & $0$    \\
$V_1^d$       & $0.23 \pm 0.03$ & $1/3$  \\
$f_1^d$       & $0.40 \pm 0.05$ & $1/3$  \\
$f_2^d$       & $0.22 \pm 0.05$ & $4/15$ \\
$f_1^u$       & $0.07 \pm 0.05$ & $1/10$ \\
\bottomrule
\end{tabular}
\end{table}

The three amplitudes $V_1$, $A_1$ and $T_1$ carry additive $x^2$
mass-correction partners of twist five,
$V_1\to V_1+\tfrac{x^2m_N^2}{4}V_1^{M}$ and likewise for $A_1$ and $T_1$, with
the same Dirac structures~\cite{Braun:2006hz}. These are included throughout.
Switching them off changes nothing in $E_T$ and $\hT$, which do not receive
contributions from the Lorentz structures these amplitudes reach, and changes
$H_T$ by $0.2\%$. They are therefore not a source of uncertainty here.

\subsection{Auxiliary parameters}
\label{sec:aux}

\begin{figure}[htb]
\centering
\includegraphics[width=0.4\textwidth]{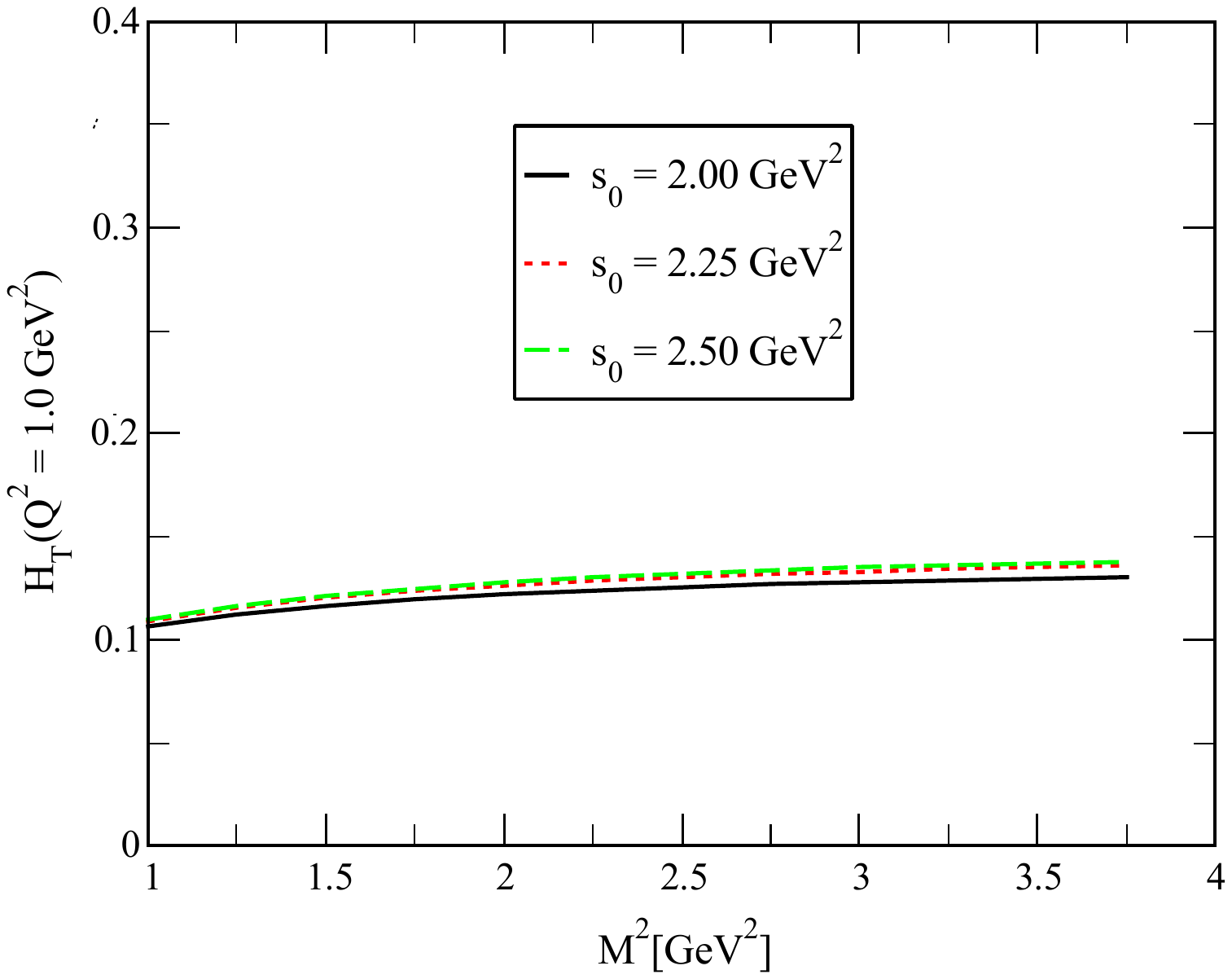}~~~~~~
\includegraphics[width=0.4\textwidth]{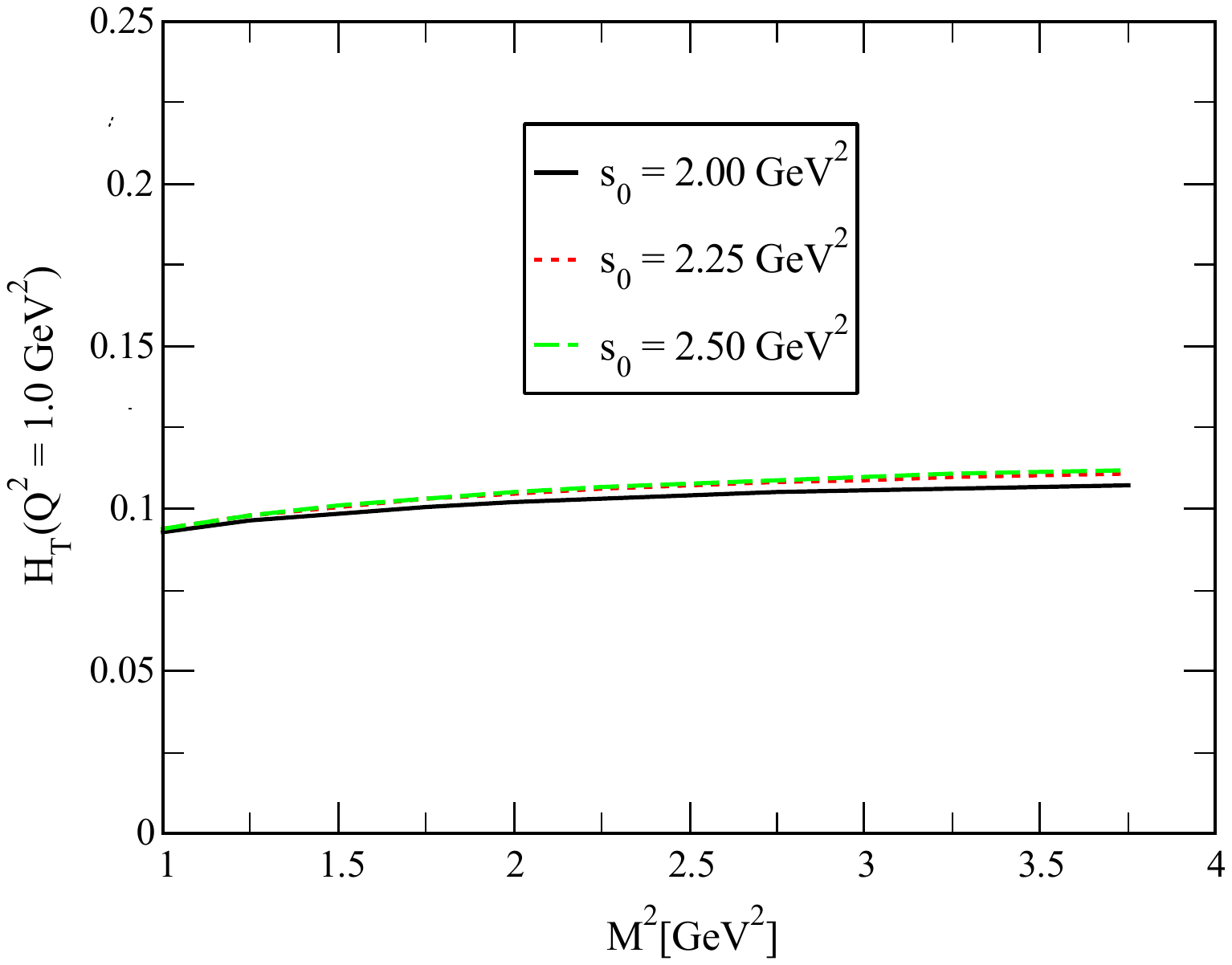}\\
\includegraphics[width=0.4\textwidth]{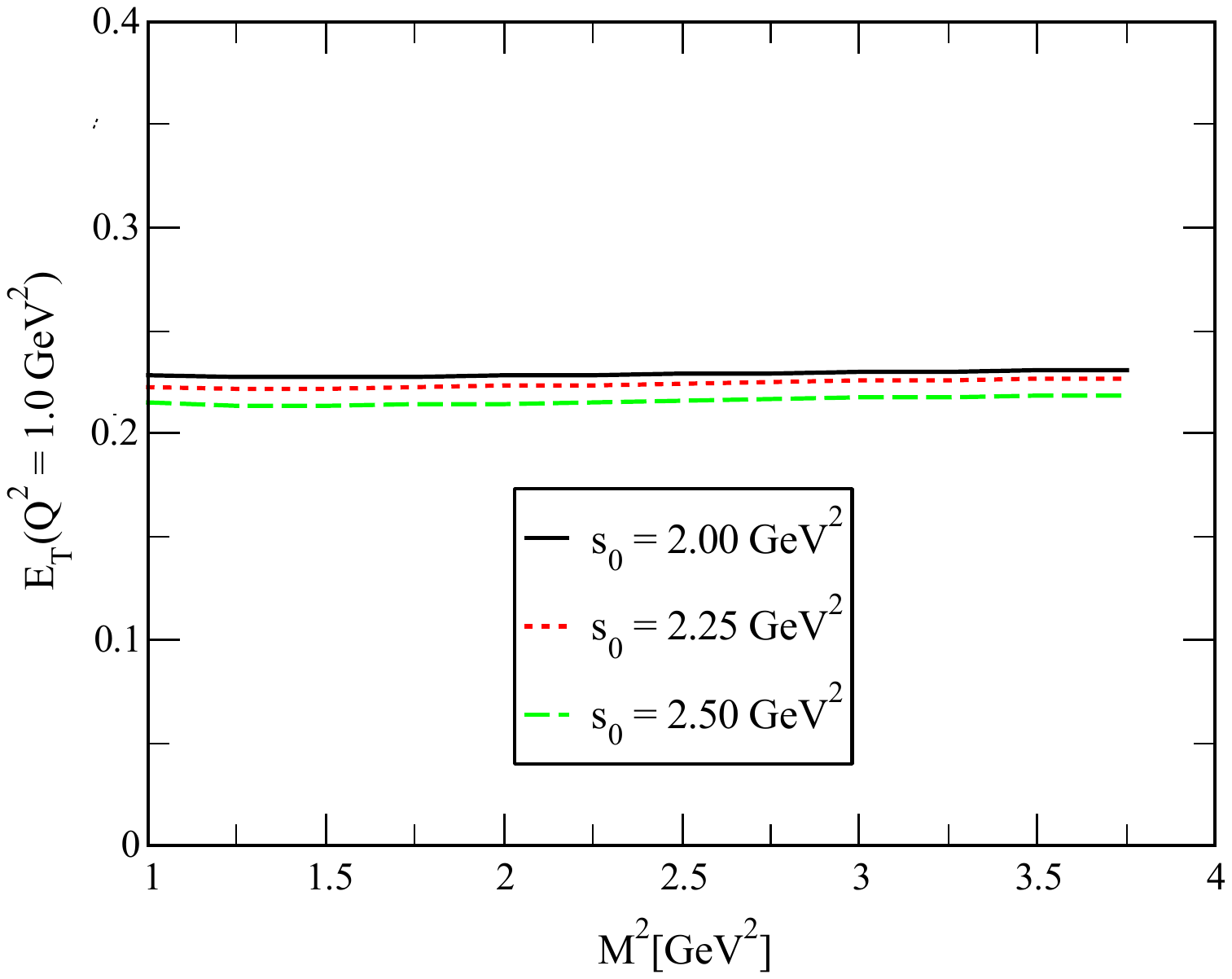}~~~~~~
\includegraphics[width=0.4\textwidth]{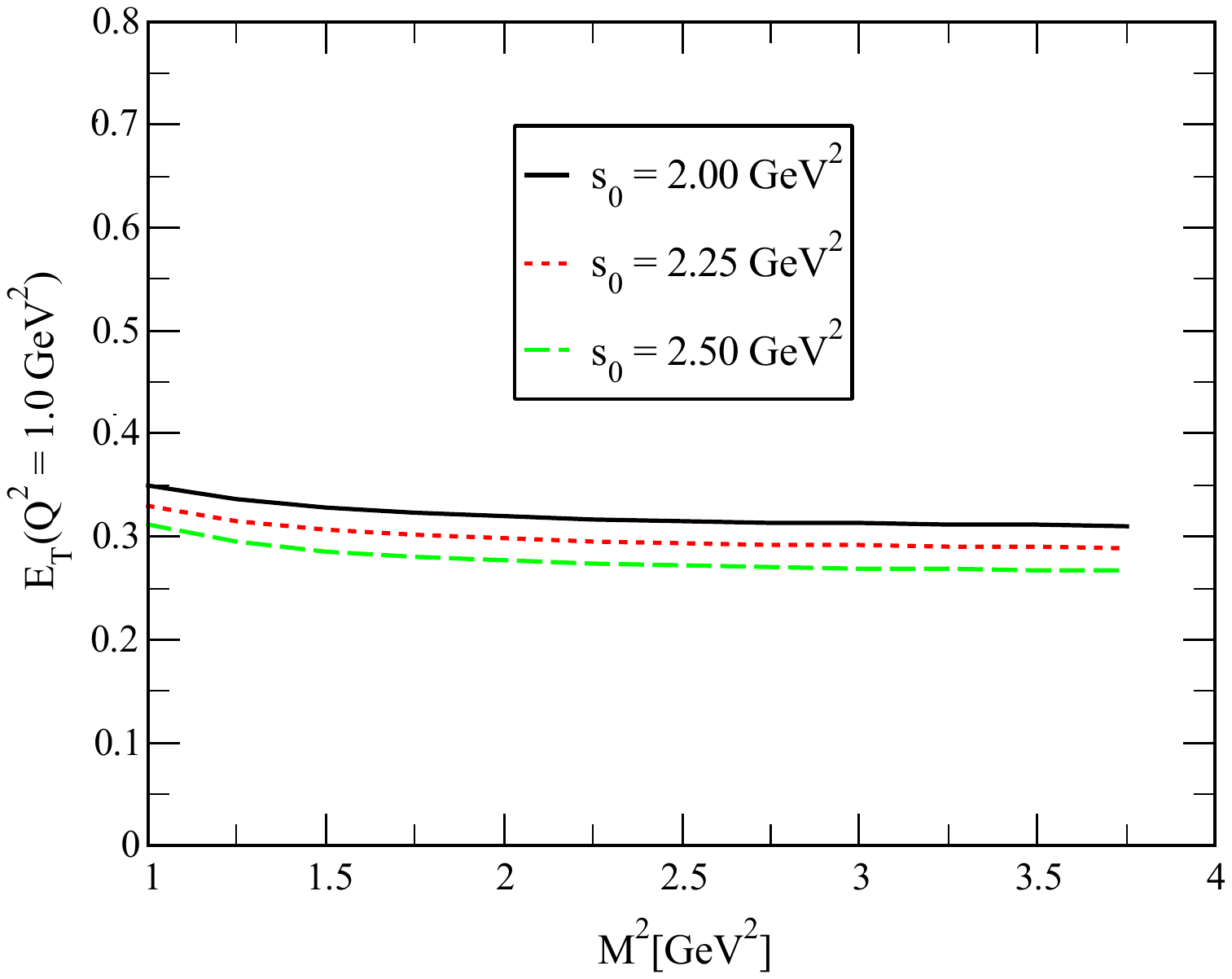}\\
\includegraphics[width=0.4\textwidth]{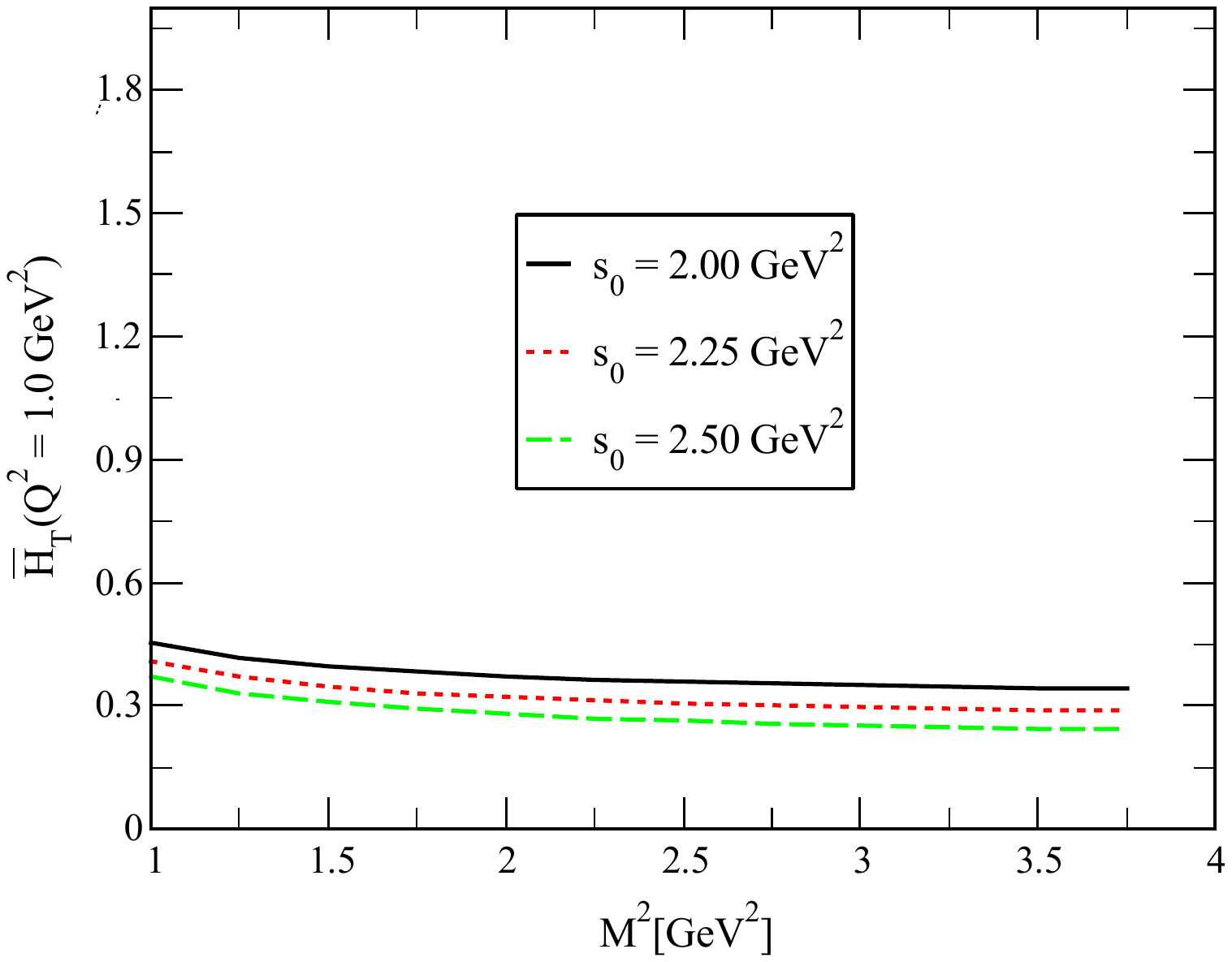}~~~~~~
\includegraphics[width=0.4\textwidth]{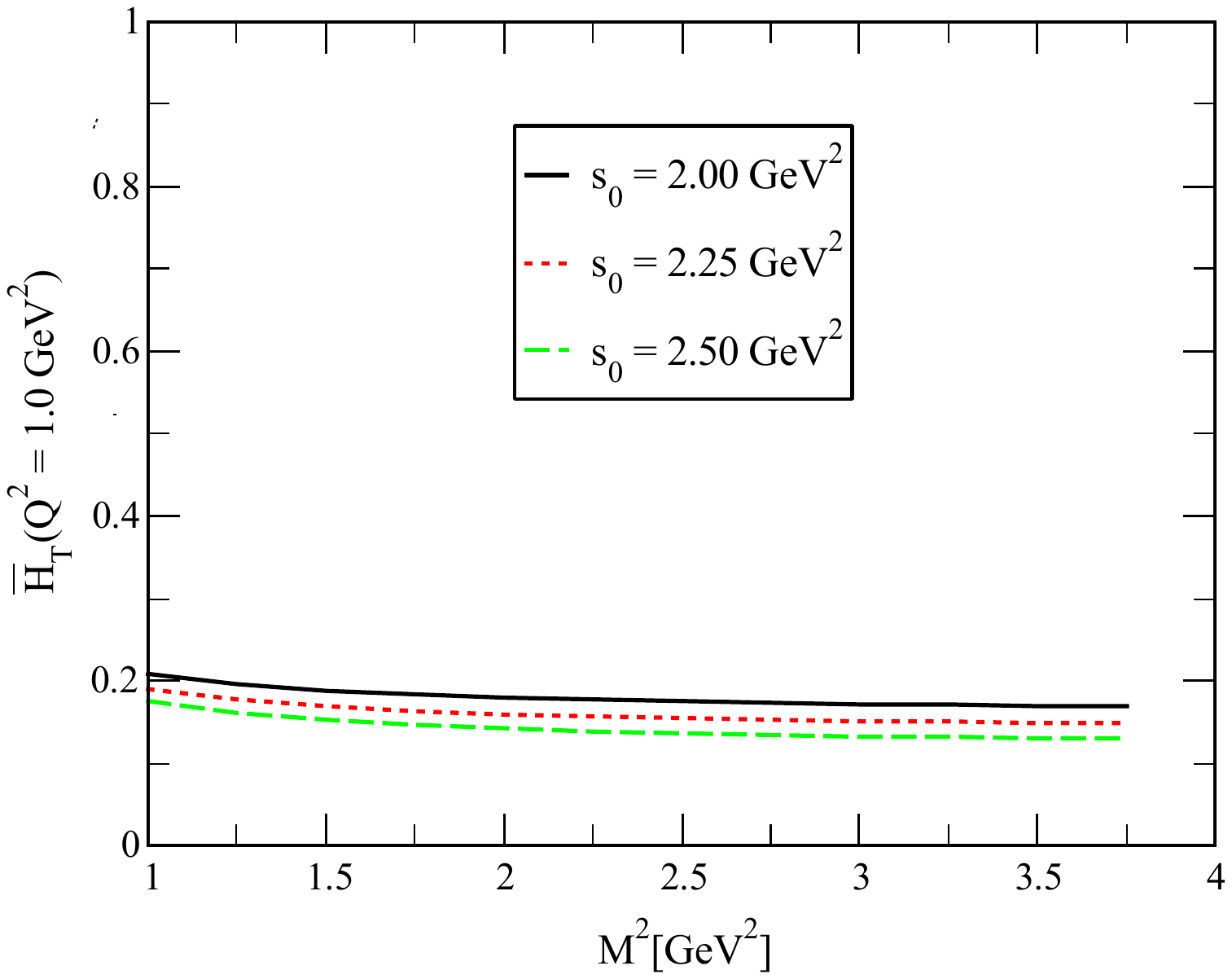}
\caption{Dependence of the isoscalar chiral-odd form factors on the Borel mass,
shown over $M^{2}\in[1.0,4.0]\GeVsq$ so that the approach to the plateau is
visible; the working window of Eq.~\eqref{eq:windows} is the subinterval
$[1.0,2.0]\GeVsq$. The curves are at $Q^{2}=1\GeVsq$ and for the three values of
the continuum threshold $s_0=2.0$, $2.25$ and $2.5\GeVsq$. Rows, from top to
bottom: $H_T$, $E_T$ and $\hT$. Left column: Set~I distribution amplitudes;
right column: Set~II. The flatness across the working window is what justifies
neglecting the Borel-mass systematic in the error budget; the spread between the
three curves is the $s_0$ uncertainty quoted in Table~\ref{tab:ff}.}
\label{fig:M2}
\end{figure}

\begin{figure}[htb]
\centering
\includegraphics[width=0.4\textwidth]{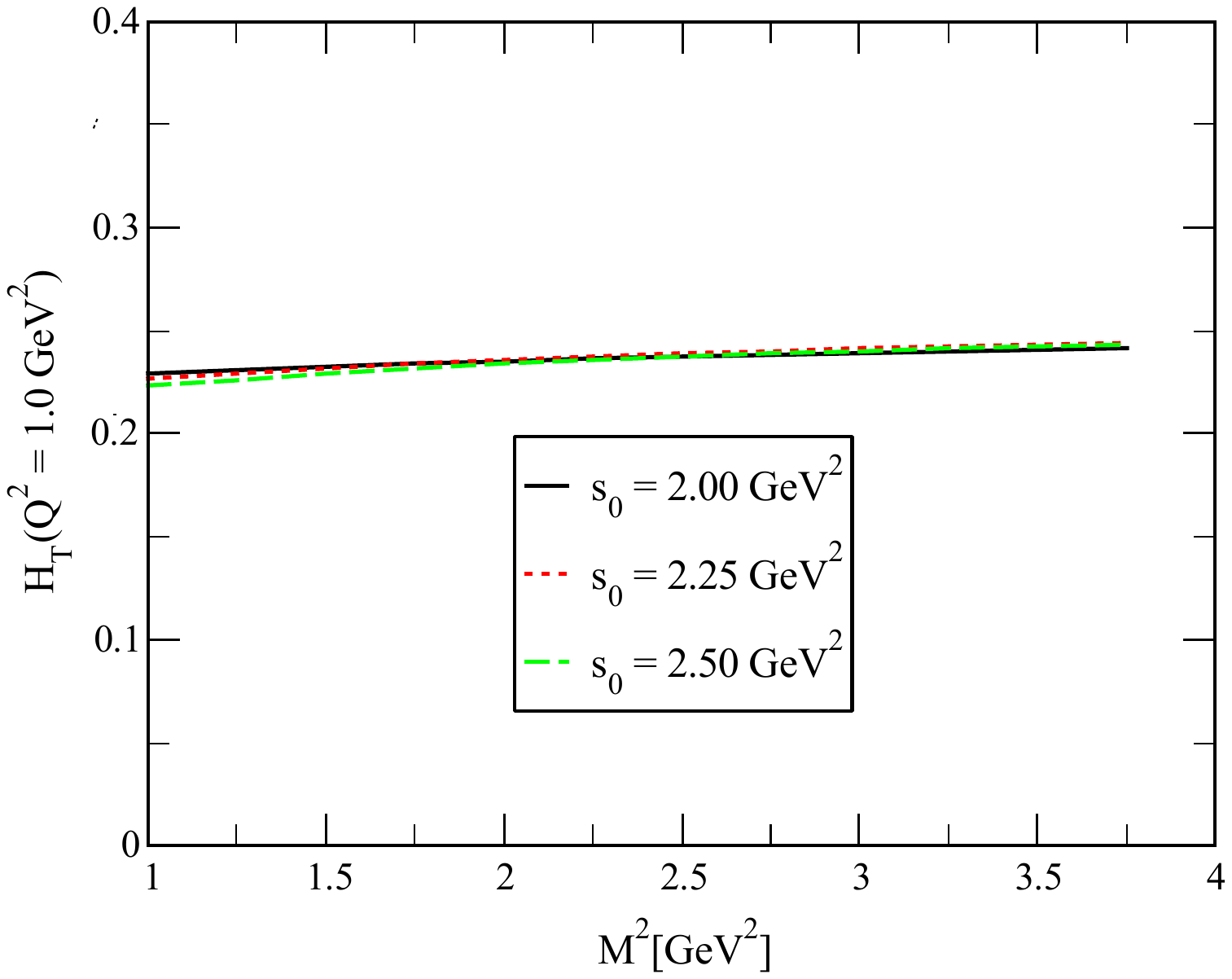}~~~~~~
\includegraphics[width=0.4\textwidth]{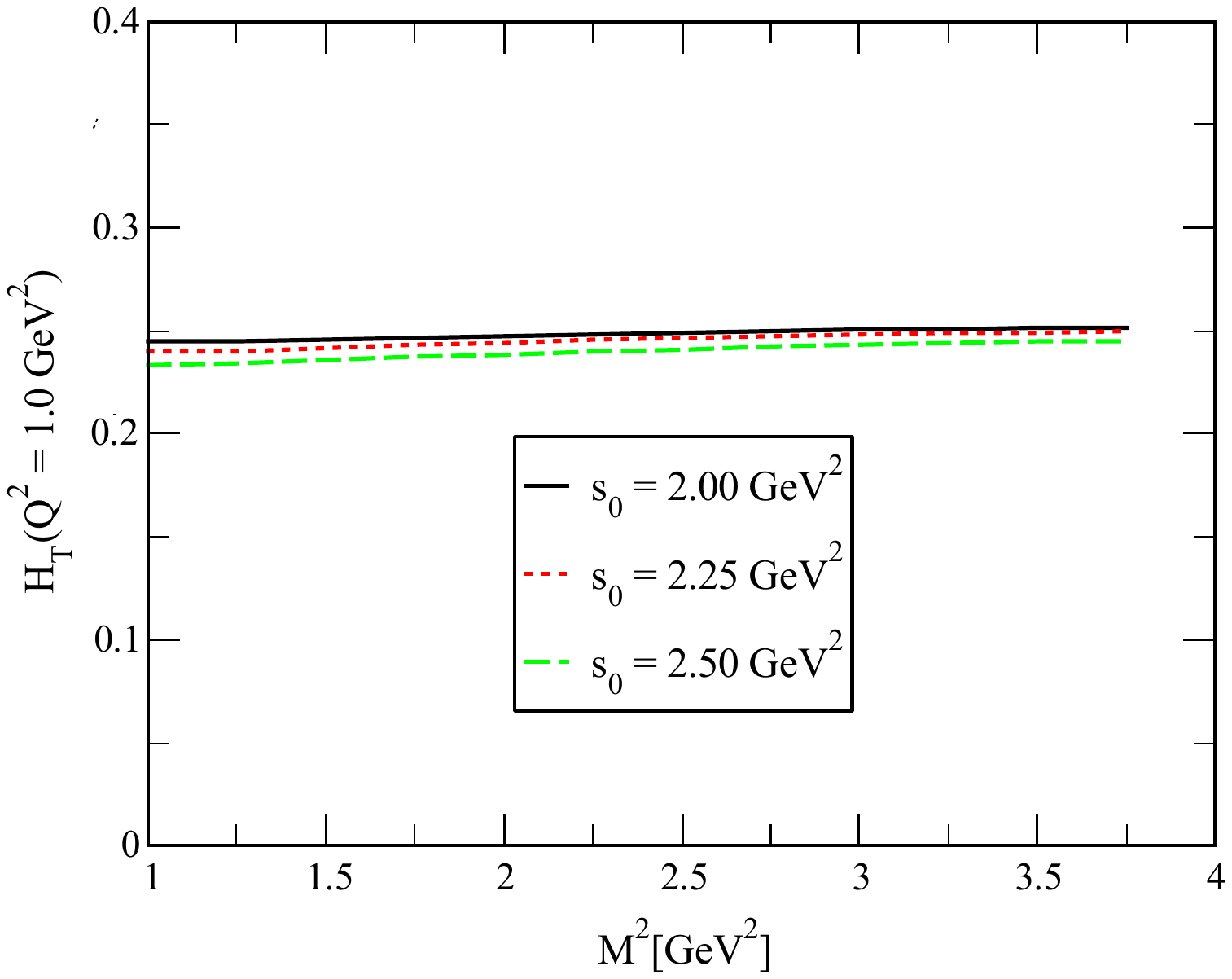}\\
\includegraphics[width=0.4\textwidth]{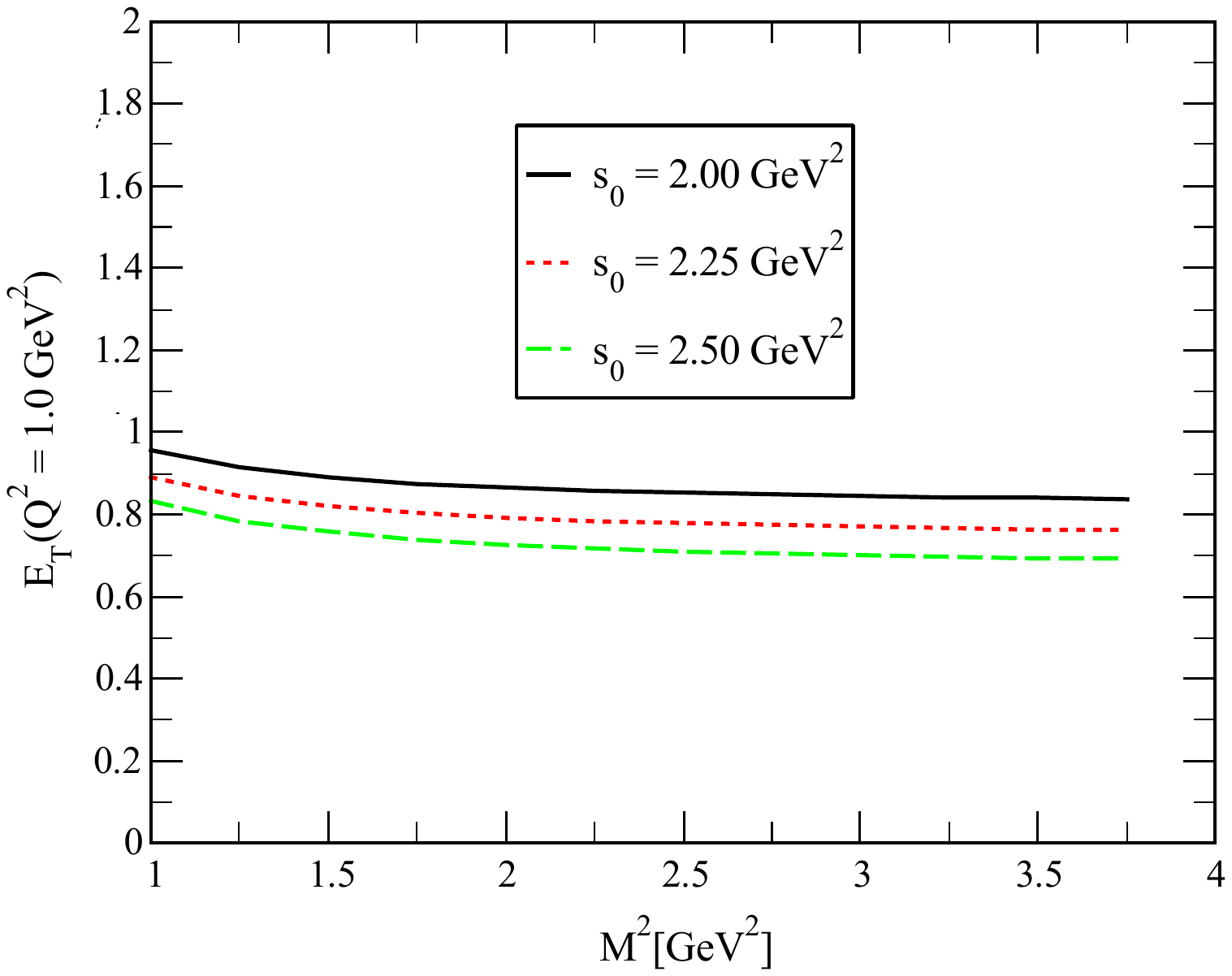}~~~~~~
\includegraphics[width=0.4\textwidth]{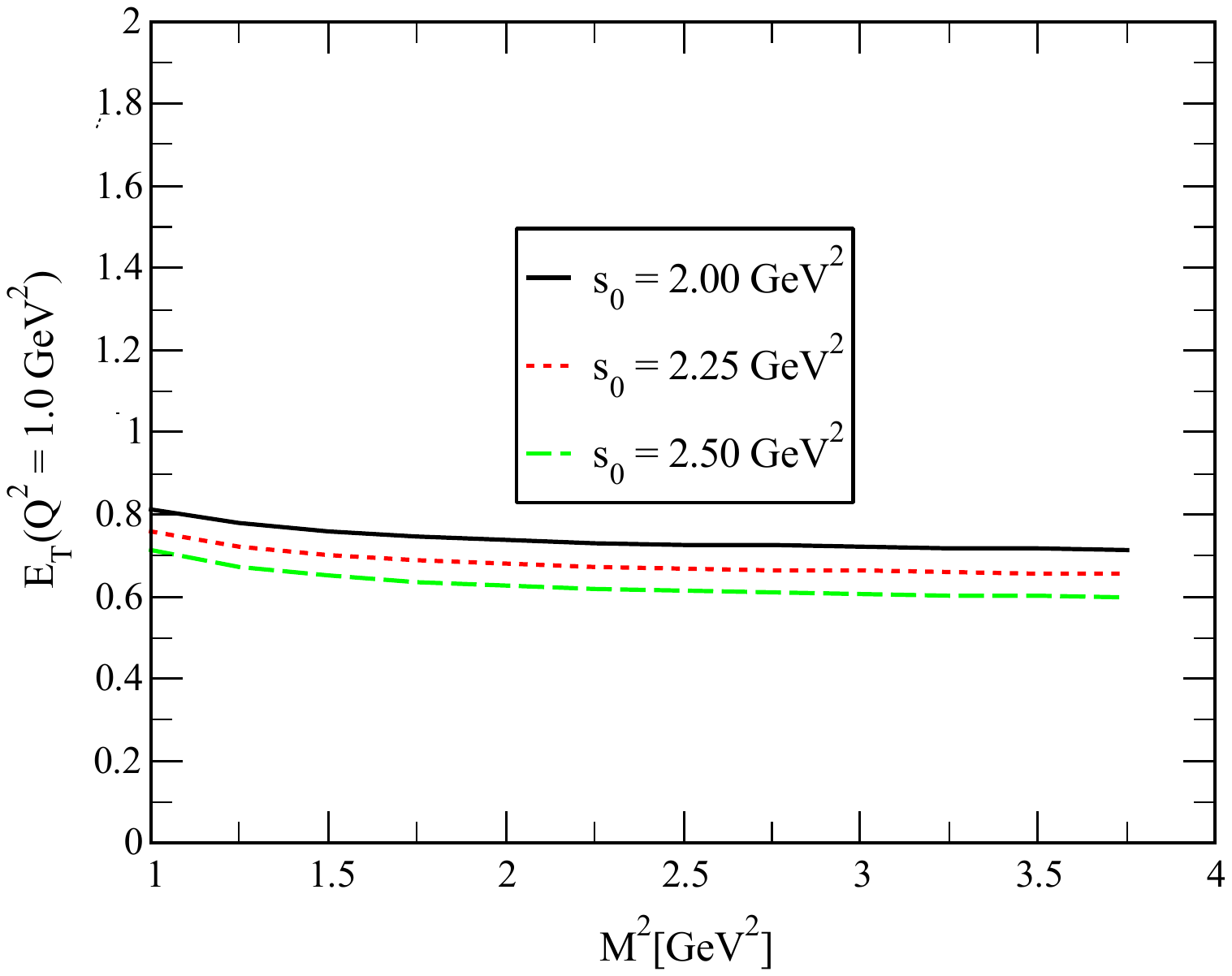}\\
\includegraphics[width=0.4\textwidth]{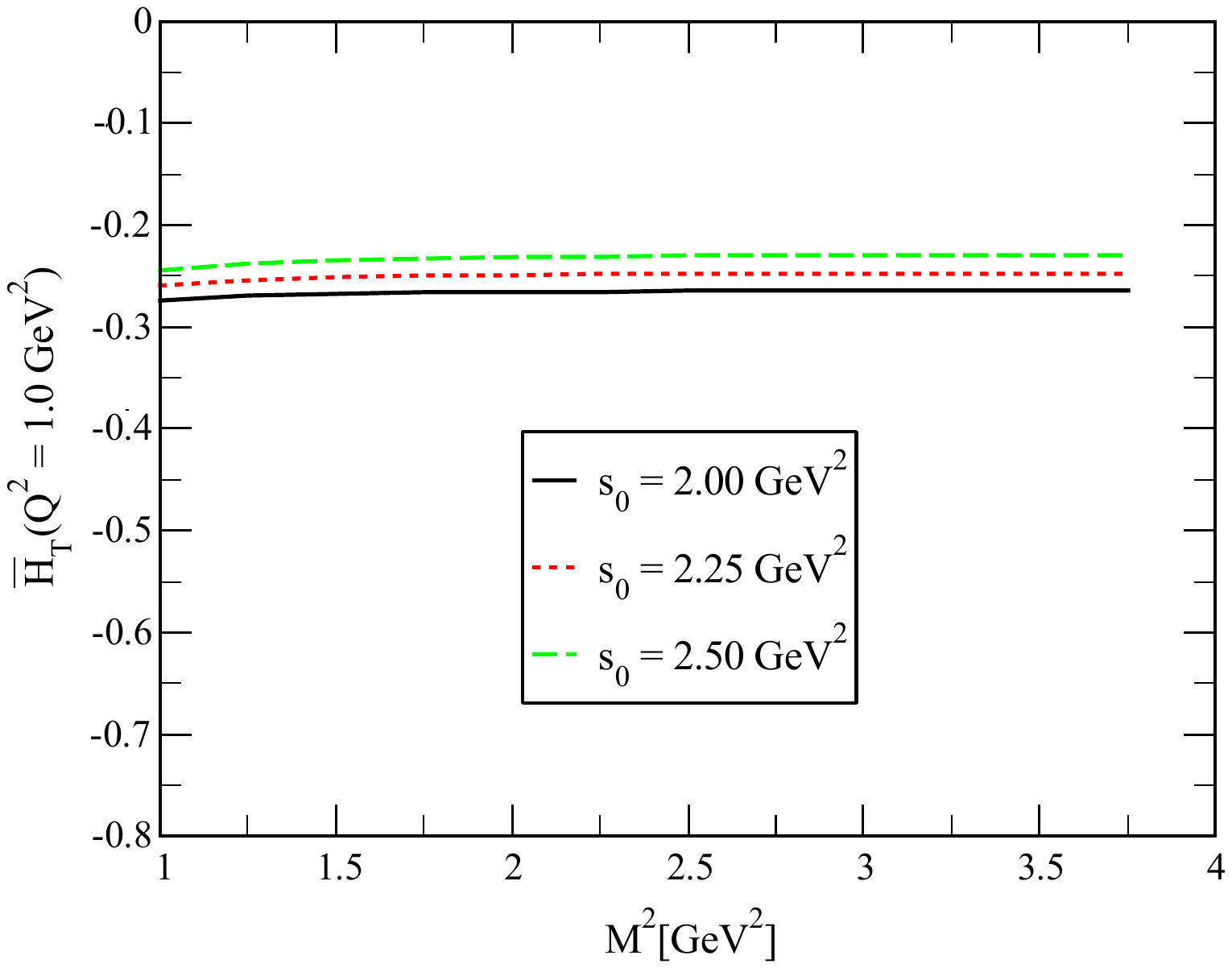}~~~~~~
\includegraphics[width=0.4\textwidth]{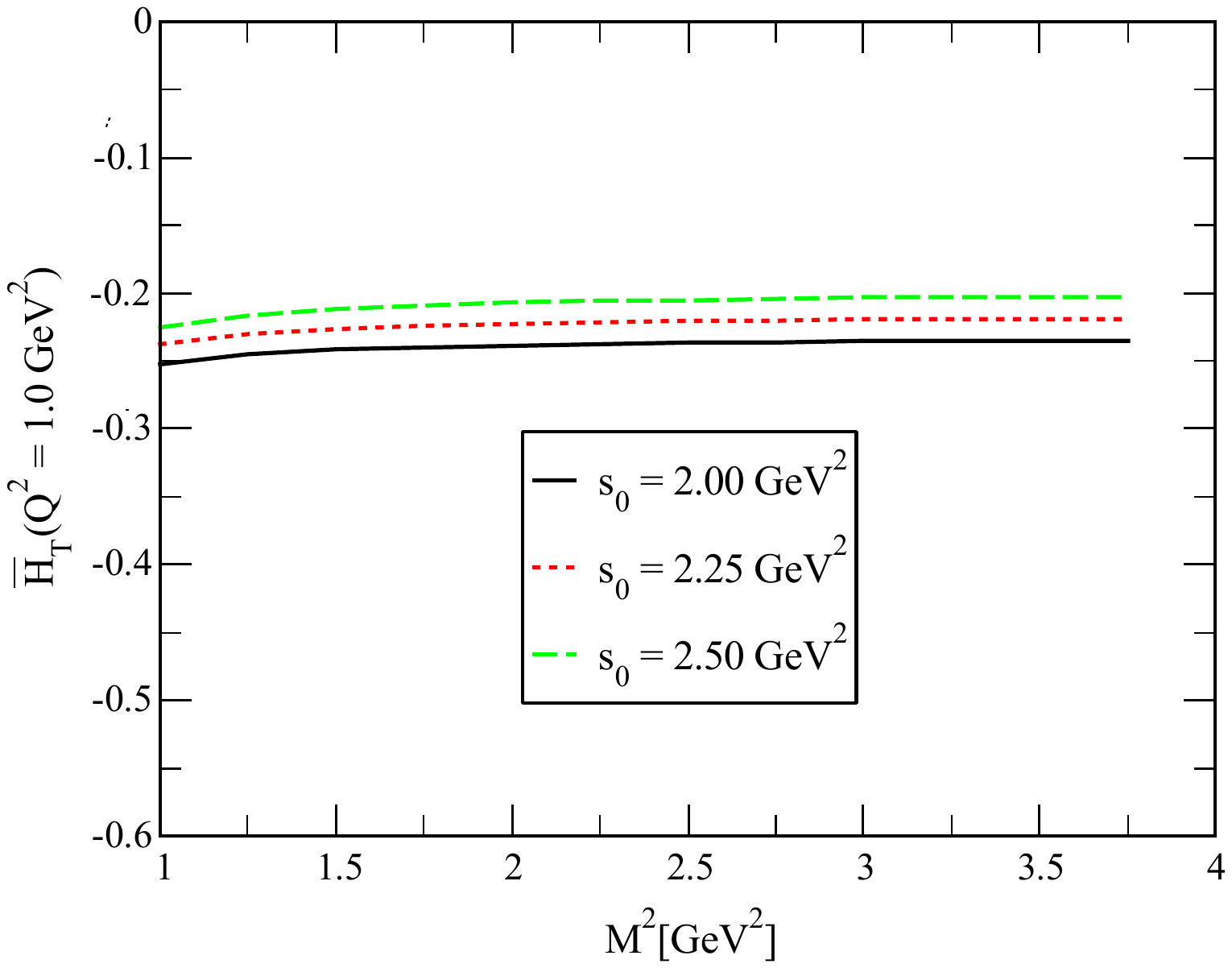}
\caption{Same as Fig.~\ref{fig:M2}, but for the isovector chiral-odd form factors.}
\label{fig:M21}
\end{figure}

The results depend on the Borel mass $M^2$ and the continuum threshold $s_0$,
and must be insensitive to both. The working regions are fixed before any form
factor is extracted, so that the criteria do not depend on the values they
produce. The upper limit of $M^2$ follows from requiring convergence of the
light-cone expansion, that is, that the highest-twist terms retained remain a
small fraction of the total; the lower limit follows from requiring that the
highest power of $1/M^2$ does not dominate, which is equivalent to demanding a
sufficiently large pole contribution. The threshold is set from the mass of the
first excitation and varied within a physically motivated interval. This gives
\begin{equation}
1.0\GeVsq\le M^2\le 2.0\GeVsq ,\qquad
2.0\GeVsq\le s_0\le 2.5\GeVsq .
\label{eq:windows}
\end{equation}
Across this window the pole contribution, defined as the ratio of the
ground-state term to the total, lies between $0.55$ and $0.74$, while the
highest-twist terms retained in the expansion enter below the $0.2\%$ level, so
both criteria are met comfortably.
The dependence on the Borel mass is shown in
Figs.~\ref{fig:M2}--\ref{fig:M21}. Within the window the results are stable to
the point that the $M^2$ variation is negligible against the other sources, and
we do not carry it in the error budget; the uncertainty is dominated by the
$s_0$ variation and by the spread between the two distribution-amplitude sets.

\subsection{Extraction of the moments}
\label{sec:extraction}

The sum rules are reliable for $Q^2\gtrsim1\GeVsq$, so the moments at $Q^2=0$
are obtained by extrapolation with
\begin{equation}
F(Q^{2})=\frac{f_0}{\big(1+Q^{2}/m^{2}\big)^{p}} ,
\label{eq:fitform}
\end{equation}
in which the exponent is left free rather than fixed to a monopole or dipole
form, so that the mass scale and the fall-off are determined separately by the
computed points rather than tied to one another by the choice of ansatz. The fit
range is fixed by the region in which the sum rules are valid and is not varied:
narrowing it discards valid data rather than probing a systematic, and the
sensitivity it would expose is already carried by the $s_0$ variation.

The extrapolation is performed separately at each of the three continuum
thresholds, and the spread of the resulting $f_0$ is quoted as the uncertainty.
Treating that spread instead as a point-by-point error would understate it,
since the three curves shift coherently rather than fluctuating independently.
The distribution-amplitude systematic is not folded into that number but
displayed by quoting both parameter sets throughout, and the $M^{2}$ dependence
is negligible as noted above.

The dipole moment requires a choice that the other two do not. Both $g_T$ and
$Q_T$ are a single form factor times a constant, so forming the combination
before or after the extrapolation is the same operation. For $\kappa_T$ it is
not: one may extrapolate $E_T$ and $\hT$ separately and add the static limits,
or form $E_T+2\hT$ pointwise in $Q^{2}$ and extrapolate that. The two target the
same number at $Q^{2}=0$ and differ only in which object the multipole ansatz is
applied to. We take the second. The combination is not an arbitrary sum: as
Eq.~\eqref{eq:hadreduced} shows, it is the coefficient of a single structure of
the ordered basis, $q_\mu p_\nu-q_\nu p_\mu$, so within this calculation it is
as much a form factor of the correlator as $E_T$ and $\hT$ are individually. A
further practical gain is that the $s_0$ dependence cancels between the two
before the fit rather than after it, which roughly halves the uncertainty.

The two routes agree. Summing the static limits of Table~\ref{tab:ff} instead
gives $\kappa_T^{u}=4.28\pm1.19$ and $\kappa_T^{d}=1.84\pm0.58$ for Set~I and
$\kappa_T^{u}=3.62\pm1.02$ and $\kappa_T^{d}=1.51\pm0.46$ for Set~II, the
uncertainties of $E_T(0)$ and $2\hT(0)$ being added in quadrature. These
uncertainties are about twice those of the combination fit because the
quadrature sum treats the two static limits as independent, whereas their $s_0$
variations are correlated and partly cancel when $E_T+2\hT$ is formed pointwise.
Compared with the values of Table~\ref{tab:2d}, the differences are about
$1\sigma$ and $0.4\sigma$ for Set~I and $0.6\sigma$ and $0.3\sigma$ for Set~II.
The agreement is closest in the $d$ channel, where the fitted combination
returns $\kappa_T^{d}=-E_T^{d}$ to four digits in both sets, exactly as
Eq.~\eqref{eq:dsector} requires; this confirms the identity of
Eq.~\eqref{eq:identity} at the level of the $Q^{2}$ dependence rather than of
the moment alone.

The flavour components are obtained directly from their own sum rules and the
isospin combinations are constructed from them, for all three moments alike.
Fitting the isospin combinations directly instead gives
$\kappa_T^{u-d}=1.31(23)$ and $\kappa_T^{u+d}=4.50(84)$ for Set~I and
$1.46(27)$ and $4.38(78)$ for Set~II, which reproduce the entries of
Table~\ref{tab:2d} to better than $0.2\sigma$ in all four cases.

An entry is quoted as a number only if the fitted shape parameter is stable
against $s_0$ and lies in a physically sensible range and the total uncertainty
is small compared with the central value; entries failing these conditions are
reported together with the reason rather than omitted.

\section{Results and discussion}
\label{sec:results}

\begin{figure}[htb]
\centering
\includegraphics[width=0.45\textwidth]{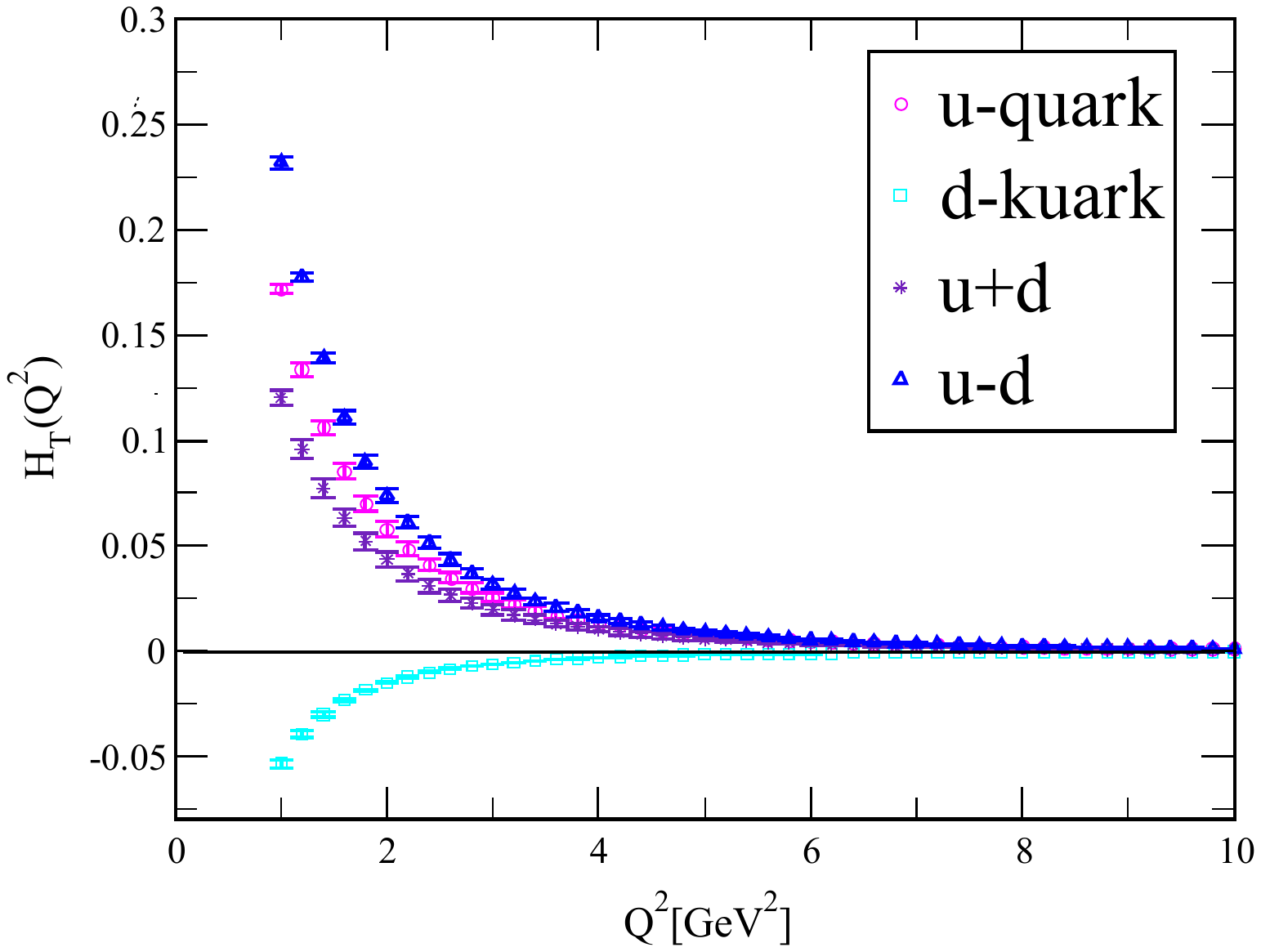}~~~~~~
\includegraphics[width=0.45\textwidth]{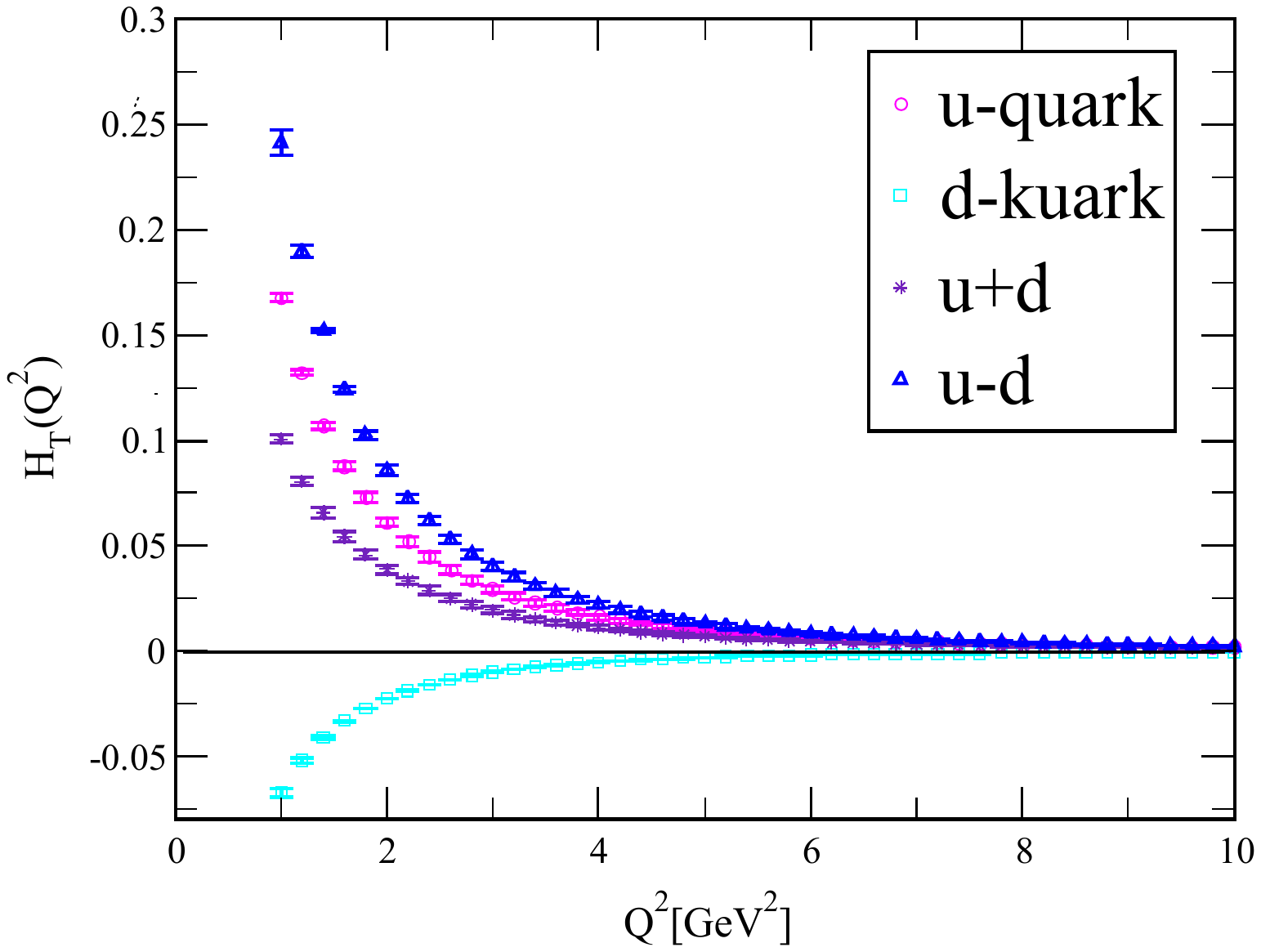}\\
\includegraphics[width=0.45\textwidth]{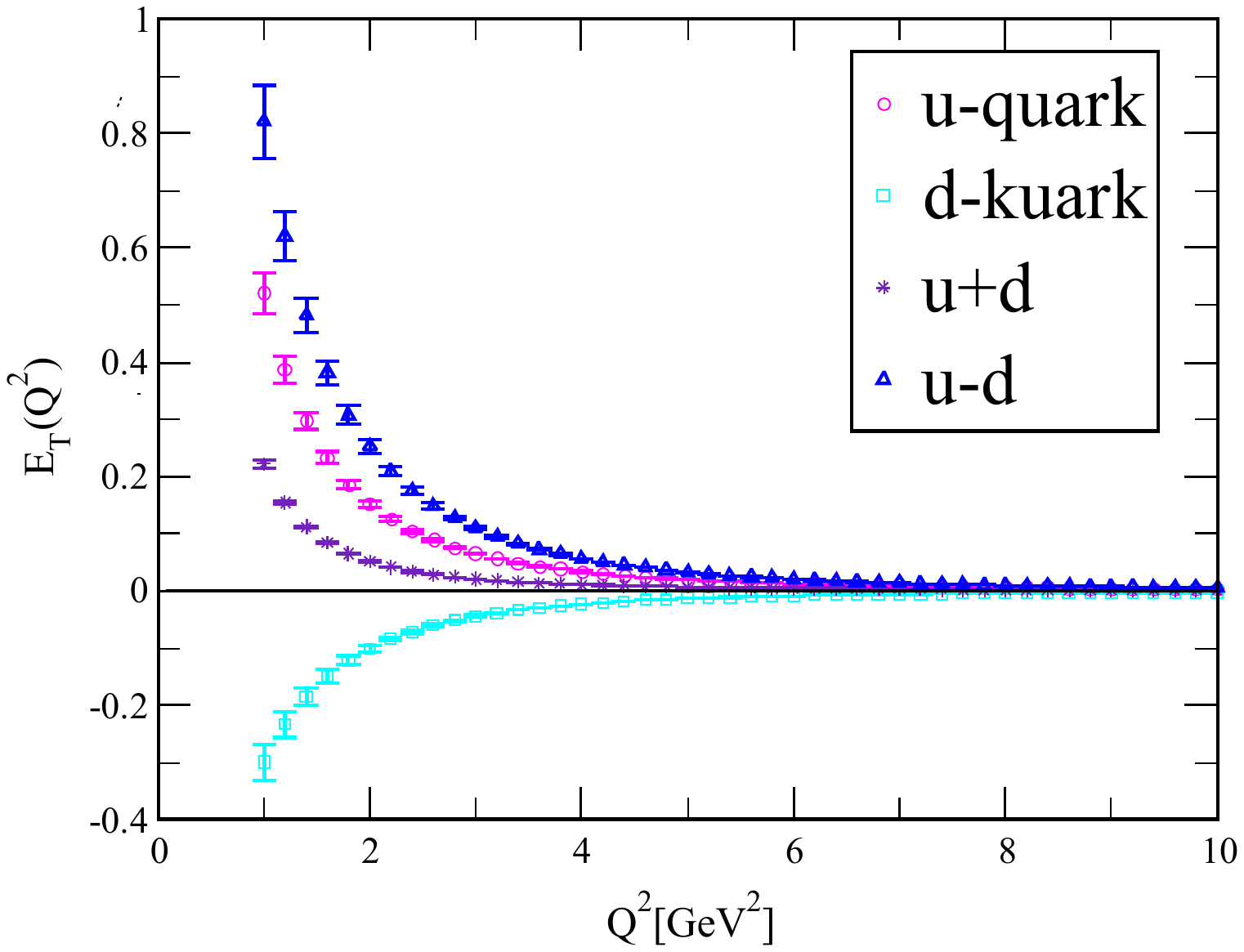}~~~~~~
\includegraphics[width=0.45\textwidth]{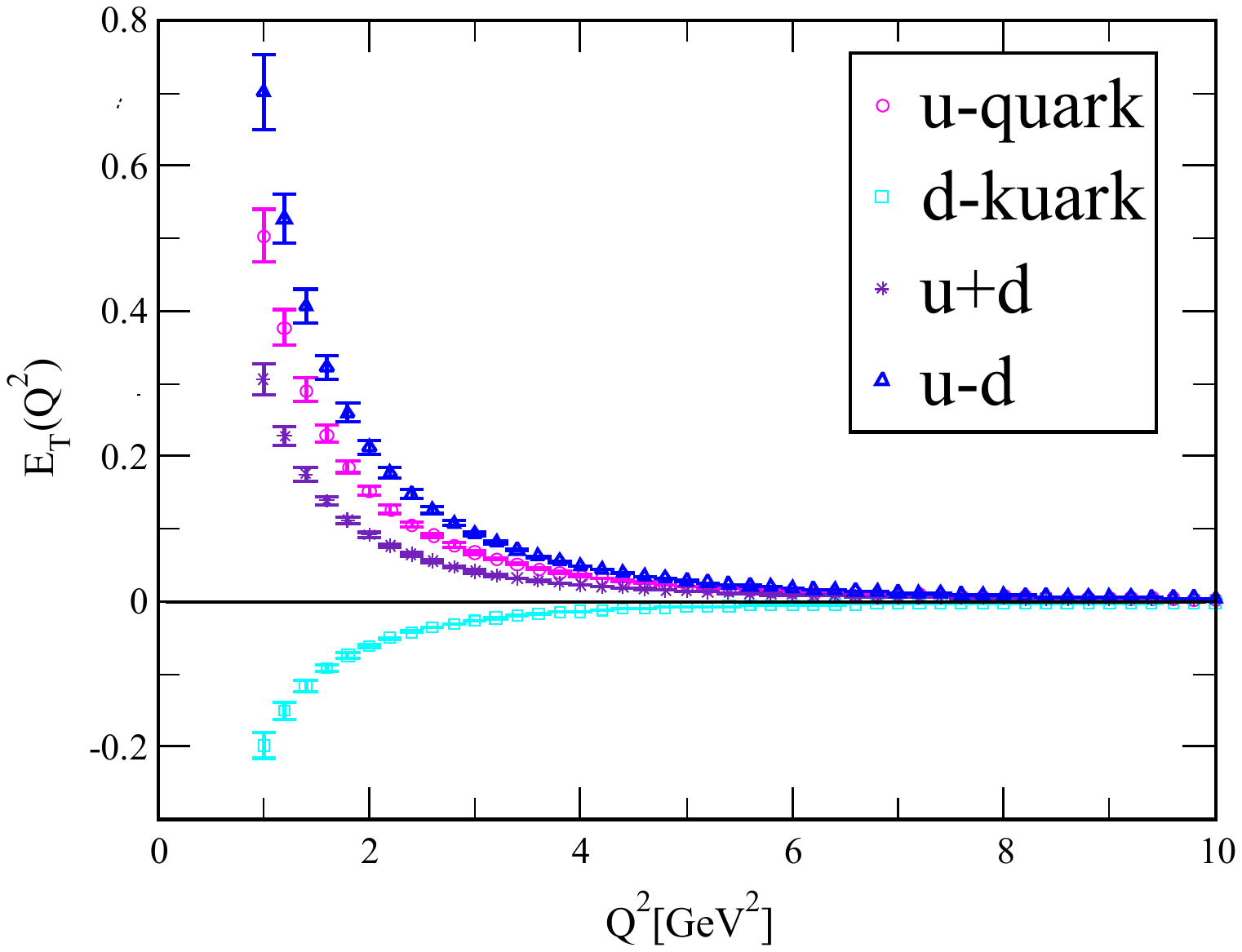}\\
\includegraphics[width=0.45\textwidth]{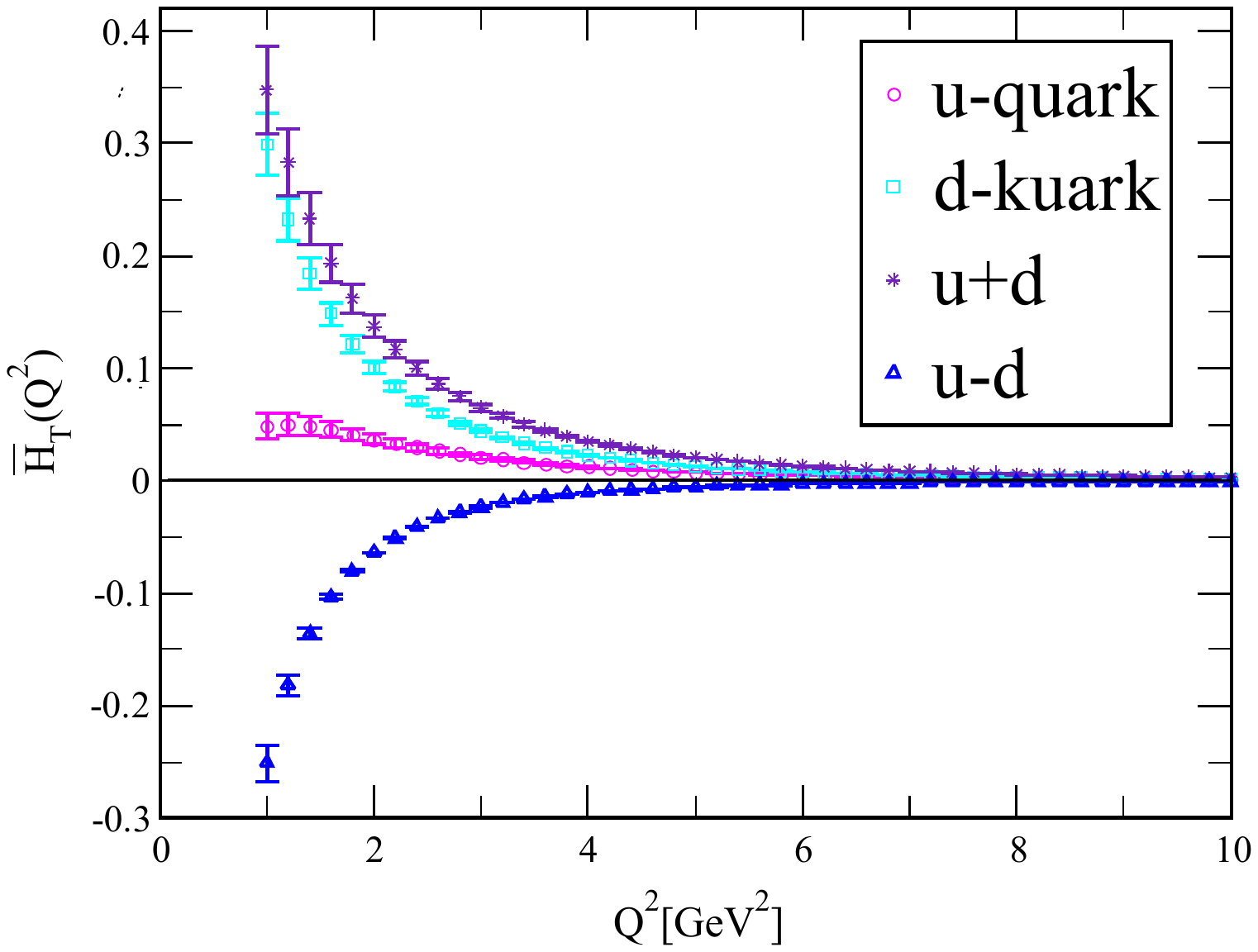}~~~~~~
\includegraphics[width=0.45\textwidth]{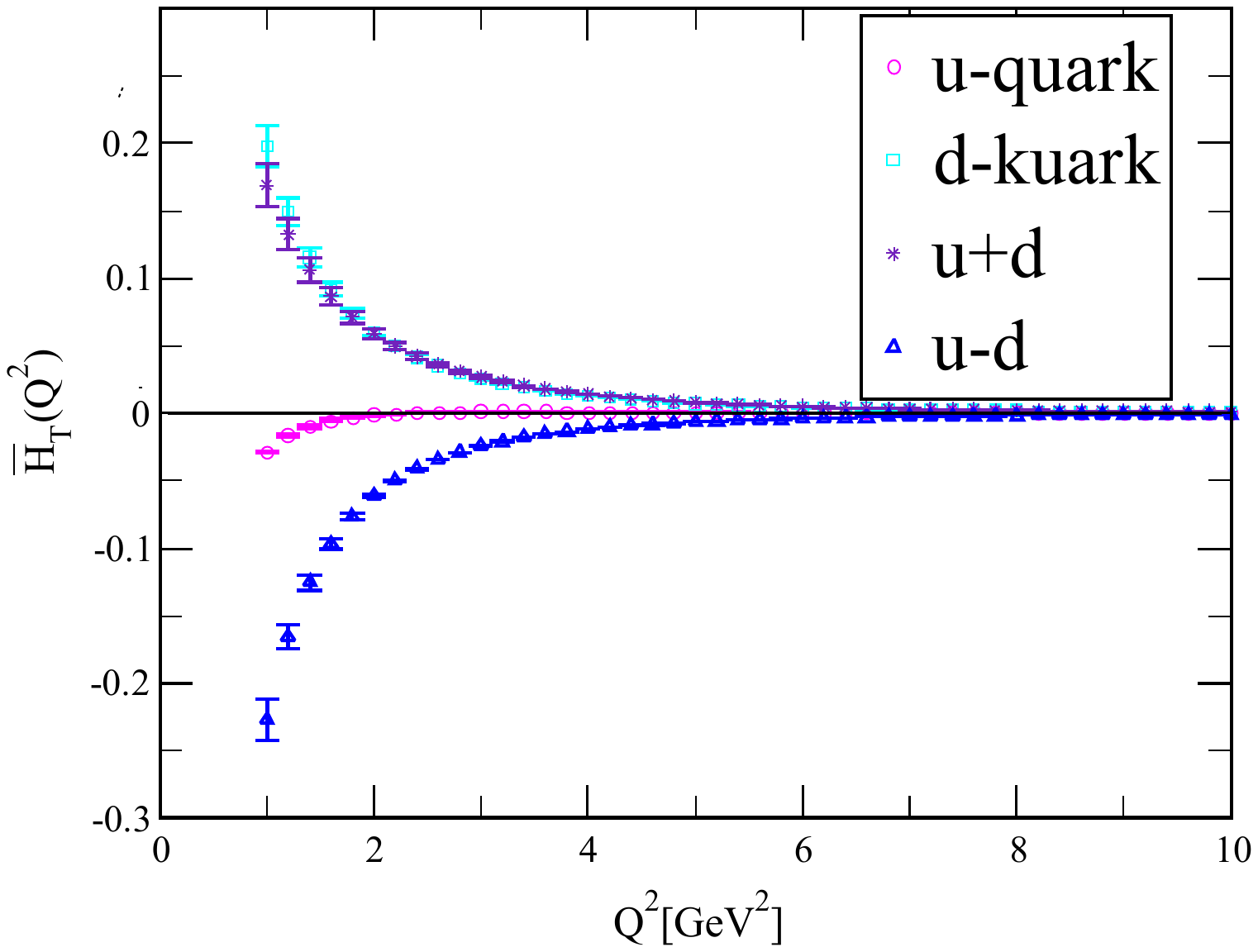}
\caption{$Q^{2}$ dependence of the chiral-odd form factors of the nucleon over
the interval $Q^{2}\in[1,10]\GeVsq$, the region in which the sum rules are
reliable. Rows, from top to bottom: $H_T$, $E_T$ and $\hT$. Left column: Set~I
distribution amplitudes; right column: Set~II. Each panel shows the $u$- and
$d$-quark contributions together with the isoscalar and isovector combinations
$u{+}d$ and $u{-}d$, with bands from the $s_0$ variation.}
\label{fig:Q2}
\end{figure}

Figure~\ref{fig:Q2} shows the $Q^{2}$ dependence of the three form factors over
the region in which the sum rules apply. These curves are the direct output of
the calculation, and the two structural results of Secs.~\ref{sec:isoscalar}
and~\ref{sec:identity} are properties of the sum rules themselves, holding at
every $Q^{2}$ in that region; the values at $Q^2=0$, which Table~\ref{tab:ff}
collects for both distribution-amplitude sets, and the moments built from them
rely in addition on the extrapolation of Sec.~\ref{sec:extraction}. For $H_T$ and
$E_T$ the flavour entries are the direct sum-rule fits and the isospin
combinations are derived from them; for $\hT$ the reverse, because the direct
$\hT^{u}$ fit does not meet the quotability conditions of
Sec.~\ref{sec:extraction} while both isospin combinations do.

That reversal makes the $d$-quark entry a check on the extrapolation itself.
Equation~\eqref{eq:identity} makes $\hT^{d}=-E_T^{d}$ exact on the QCD side, so
reconstructing $\hT^{d}$ from the two isospin fits ought to return $1.56$ and
$1.37$; it returns $1.70$ and $1.44$, a departure of $9\%$ and $5\%$. This is
the price of extrapolating the two isospin combinations independently rather
than the $d$ channel itself, and it sets the size of the extraction systematic
carried by every $\hT$ and $Q_T$ entry in Tables~\ref{tab:ff}
and~\ref{tab:2d}. It is also why $Q_T^{d}$ in Table~\ref{tab:2d} is $3.40$
rather than the $2\kappa_T^{d}=3.12$ that the identity would give:
$\kappa_T^{d}$ comes from the combination fit of Sec.~\ref{sec:extraction} and
therefore respects the identity exactly, while $Q_T^{d}=2\hT^{d}(0)$ inherits
the reconstruction systematic just quoted.

\begin{table}[htp]
\caption{Chiral-odd form factors at $Q^2=0$ and $\mu^2=1\GeVsq$.}
\label{tab:ff}
\begin{ruledtabular}
\begin{tabular}{llcccc}
 & & $u$ & $d$ & $u{+}d$ & $u{-}d$\\
\hline
 Set I & $H_T(0)$     & $~~0.83\pm0.06$ & $-0.44\pm0.03$ & ~~$0.39\pm0.07$ & $~~1.27\pm0.07$\\
      & $E_T(0)$     & $~~5.28\pm1.08$ & $-1.56\pm0.29$ & $~~3.72\pm1.12$ & $~~6.84\pm1.12$\\
      & $\hT(0)$     & $-0.50\pm0.25$ & $~~1.70\pm0.25$ & $~~1.20\pm0.26$ & $-2.20\pm0.42$\\
\hline
 Set II & $H_T(0)$     & $~~0.73\pm0.04$ & $-0.35\pm0.04$ & $~~0.38\pm0.06$ & $~~1.08\pm0.06$\\
      & $E_T(0)$     & $~~4.94\pm0.94$ & $-1.37\pm0.23$ & $~~3.57\pm0.97$ & $~~6.31\pm0.97$\\
      & $\hT(0)$     & $-0.66\pm0.20$ & $~~1.44\pm0.20$ & $~~0.78\pm0.15$ & $-2.10\pm0.36$\\
\end{tabular}
\end{ruledtabular}
\end{table}

\begin{table}[htp]
\caption{Tensor multipole moments, Eq.~\eqref{eq:2d}. The $\kappa_T$ entries
come from the combination fit of Sec.~\ref{sec:extraction}, the $Q_T$ entries
from $2\hT(0)$ of Table~\ref{tab:ff}; the two therefore differ in the $d$
channel by the reconstruction systematic discussed in the text.}
\label{tab:2d}
\begin{ruledtabular}
\begin{tabular}{llcccc}
 & & $u$ & $d$ & $u{+}d$ & $u{-}d$\\
\hline
 Set I & $g_T$        & $~~0.83\pm0.06$ & $-0.44\pm0.03$ & $~~0.39\pm0.07$ & $~~1.27\pm0.07$\\
      & $\kappa_T$   & $~~2.91\pm0.54$ & $~~1.56\pm0.30$ & $~~4.47\pm0.62$ & $~~1.35\pm0.62$\\
      & $Q_T$        & $-1.00\pm0.49$ & $~~3.40\pm0.49$ & $~~2.41\pm0.53$ & $-4.40\pm0.83$\\
\hline
 Set II & $g_T$        & $~~0.73\pm0.04$ & $-0.35\pm0.04$ & $~~0.38\pm0.06$ & $~~1.08\pm0.06$\\
      & $\kappa_T$   & $~~2.95\pm0.54$ & $~~1.37\pm0.23$ & $~~4.32\pm0.59$ & $~~1.58\pm0.59$\\
      & $Q_T$        & $-1.32\pm0.39$ & $~~2.88\pm0.39$ & $~~1.56\pm0.30$ & $-4.20\pm0.73$\\
\end{tabular}
\end{ruledtabular}
\end{table}

\begin{table*}[htp]
\caption{Comparison with other determinations. Every entry, including its
renormalisation scale, is taken from the original reference rather than from a
compilation. The column $\mu^2$ gives that scale in GeV$^2$; every row is then
brought to $\mu^2=1\GeVsq$ with Eq.~\eqref{eq:evol}, errors scaling with the
same factor. The factors are $0.871$ from $\mu^2=0.35$, $0.876$ from $0.36$,
$0.925$ from $0.49$ and $1.085$ from $4.0$. Lattice results are in the $\overline{\rm MS}$ scheme at
$2$ GeV. For Refs.~\cite{Ghim:2025gqo,Ghim:2026qnp} the scale is not displayed in
the expressions but is fixed by the inverse average instanton size,
$\bar\rho^{-1}=0.59$ GeV; for Ref.~\cite{Tezgin:2024tfh} the initial scale is
$\mu_0=0.7$ GeV, and the range $0.5$--$0.9$ GeV explored there moves the
evolution factor by up to $15\%$ downward and $6\%$ upward, so that row carries
an extra normalisation uncertainty of that size. For Ref.~\cite{Erkol:2011iw}
the range spans the three fit regions quoted there, $Q^{2}\in[2,10]$,
$[1.5,10]$ and $[1,10]\GeVsq$. Where a reference resolves the flavours,
the isospin combinations are constructed from them. Errors are in the last
digits; $\ldots$ means the quantity is not given in that reference.}
\label{tab:comp}
\begin{ruledtabular}
\begin{tabular}{llcccccc}
 & $\mu^2$ & $g_T^{u-d}$ & $g_T^{u+d}$ & $\kappa_T^{u-d}$ & $\kappa_T^{u+d}$
 & $Q_T^{u-d}$ & $Q_T^{u+d}$\\
\hline
 This work, Set~I                                         & $1.0$ & $1.27(7)$ & $0.39(7)$ & $1.35(62)$ & $4.47(62)$ & $-4.40(83)$ & $2.41(53)$\\
 This work, Set~II                                        & $1.0$ & $1.08(6)$ & $0.38(6)$ & $1.58(59)$ & $4.32(59)$ & $-4.20(73)$ & $1.56(30)$\\
\hline
 LCSR~\cite{Erkol:2011iw}                                 & $1.0$ & $1.15$--$2.11$ & \ldots & $1.82$--$3.86$ & \ldots & $0.86$--$1.94$ & \ldots\\
 QCD sum rules~\cite{He:1994gz}                           & $1.0$ & $1.00(71)$ & $1.00(71)$ & \ldots & \ldots & \ldots & \ldots\\
 Lattice QCD~\cite{Alexandrou:2022dtc,Alexandrou:2019brg,Gupta:2018lvp,Bhattacharya:2015wna} & $4.0$ & $1.00$--$1.11$ & $0.59$--$0.61$ & $1.14(10)$ & \ldots & \ldots & \ldots\\
 Global fit, no lattice~\cite{Cocuzza:2023oam}            & $4.0$ & $0.56(13)$ & $0.52(16)$ & \ldots & \ldots & \ldots & \ldots\\
 Global fit, with lattice~\cite{Cocuzza:2023oam}          & $4.0$ & $0.99(2)$ & $0.55(2)$ & \ldots & \ldots & \ldots & \ldots\\
 $\chi$QSM SU(3)~\cite{Ledwig:2010tu,Ledwig:2010zq}       & $0.36$ & $1.23$ & $0.67$ & $1.52$ & $4.72$ & \ldots & \ldots\\
 $\chi$QSM SU(2)~\cite{Ledwig:2010zq}                     & $0.36$ & $1.25$ & $0.66$ & $1.66$ & $4.86$ & \ldots & \ldots\\
 $\chi$QSM, LO~\cite{Ghim:2025gqo}                        & $0.35$ & $0.86$ & \ldots & \ldots & $6.63$ & $-6.11$ & \ldots\\
 $\chi$QSM, $1/N_c$~\cite{Ghim:2026qnp}                   & $0.35$ & \ldots & $0.71$ & $1.72$ & \ldots & \ldots & \ldots\\
 MIT bag model~\cite{Tezgin:2024tfh}                      & $0.49$ & $1.28$ & $0.76$ & $0.85$ & $5.61$ & \ldots & \ldots\\
 Quark-model average~\cite{Tezgin:2024tfh}                & $4.0$ & $1.20(24)$ & \ldots & $1.17(31)$ & \ldots & \ldots & \ldots\\
\end{tabular}
\end{ruledtabular}
\end{table*}

Table~\ref{tab:comp} places the results next to the other determinations, all at
a common scale. For the isovector tensor charge most of the other determinations
cluster tightly. Four independent lattice calculations give $1.00$--$1.11$ once
brought to a common scale, the external-field QCD sum rule of
Ref.~\cite{He:1994gz} gives $1.00(71)$ at the scale we work at, and the
quark-model average of Ref.~\cite{Tezgin:2024tfh} gives $1.20(24)$. The earlier
light-cone QCD sum rule is the exception: it spans $1.15$--$2.11$ across the three
fit regions quoted there, a factor of two, which already indicates how sensitive
the extrapolation to $Q^2=0$ is in this channel. Set~II, at $1.08(6)$, falls
inside the lattice band, and Set~I, at $1.27(7)$, lies just above it while
agreeing with the quark-model average; both sit in the lower part of the earlier
sum-rule range. The global fit of Ref.~\cite{Cocuzza:2023oam} gives $0.56(13)$
from data alone and $0.99(2)$ once lattice input is imposed as a prior, so the
phenomenological determination is the one entry lying clearly below the lattice
band; that tension is between data and lattice rather than with the present
work.

The isoscalar tensor charge is where we are clearly low. Our $0.38$--$0.39$ is
to be compared with $0.59$--$0.61$ from two lattice calculations and $0.55(2)$
from the lattice-constrained global fit, three determinations that agree among
themselves; the same shows up in the scale-independent ratio
$|\delta d/\delta u|$, for which we obtain $0.53$ and $0.48$ against $0.30$ in
the chiral quark-soliton model, $0.25$ in the bag model and $0.28$ in the
lattice-constrained global fit. These are one statement, not two, and the reason
is structural: as shown in Sec.~\ref{sec:isoscalar}, the isoscalar tensor charge
receives no leading-twist contribution at the level of the least suppressed
integral.

The contrast between the two isovector columns is worth noting on its own. The
tensor charge has been computed four times on the lattice with a total spread of
about $10\%$ and is further constrained by a global fit to data, whereas the
anomalous tensor magnetic moment has been computed on the lattice once, has no
phenomenological determination at all, and the theoretical values for it range
over a factor of four and a half.

The isovector dipole moment shows the widest spread of any entry in the table:
$0.85$ in the bag model, $1.14$ on the lattice, $1.17$ for the quark-model
average, $1.52$ and $1.66$ in the two soliton calculations, $1.72$ in the
mean-field calculation with $1/N_c$ corrections, $1.82$--$3.86$ across the three
fit regions of the earlier sum-rule work, and $1.35$--$1.58$ here. Our values
sit in the middle of that spread, between the lattice and quark-model
determinations on the one side and the soliton results on the other. The reason
the spread is so wide is structural rather than numerical: as discussed in
Sec.~\ref{sec:largeNc}, $\kappa_T^{u-d}$ is the residual of a cancellation
between two much larger quantities, enforced to two powers of $N_c$, so it
inherits both their uncertainties and is intrinsically the least
well-determined of the three moments in any approach. The width of the
literature spread is itself evidence of that.

For the isoscalar dipole our $4.32$--$4.47$ falls just below the two soliton
values, $4.72$ and $4.86$, and well below the bag model. For $\hT^{u-d}$ and
$Q_T^{u-d}$ we agree in sign and in order of magnitude with the mean-field
calculation while disagreeing in sign with the only other sum-rule
determination. That determination also violates the large-$N_c$ relation of
Sec.~\ref{sec:largeNc} in sign rather than in magnitude. The extraction basis is
one difference to bear in mind here: there $H_T$ and $E_T$ are read from
structures which, as shown in Sec.~\ref{sec:lcsr}, are not part of the
independent set once the Dirac strings are ordered. In the flavour decomposition
both components of the dipole moment come out positive, $\kappa_T^{u}=2.91(54)$
and $\kappa_T^{d}=1.56(30)$ for Set~I, as they do on the lattice and in the
soliton models; this is the sign pattern required by the connection between the
transverse distortion and the Boer--Mulders function~\cite{Burkardt:2005hp}.

The comparison with the soliton calculations also raises the role of sea
quarks. In the chiral quark-soliton model the matrix elements are the sum of a
contribution from the discrete valence level and one from the polarised Dirac
continuum, and the latter has been found to be important for the tensor dipole
structure~\cite{Ghim:2026qnp}. Being odd under charge conjugation, the tensor
current measures the difference between quark and antiquark contributions, so
that quark--antiquark pairs from the sea enter only to the extent that the quark
and the antiquark differ in transverse polarisation or in transverse position.
The separation of the soliton model into valence level and Dirac continuum is
moreover not the same as the partonic separation into valence and sea quarks,
the continuum contributing to antiquark as well as to quark distributions. On
the side of the method, the nucleon enters the sum rules only through its
three-quark distribution amplitudes up to twist six, and Fock components with
an additional quark--antiquark pair are not included, so a valence--sea
decomposition analogous to that of the soliton model has no counterpart here.
Whether the residual difference between our dipole moments and the soliton
values reflects such components, the different treatment of $N_c$, or the
truncation of the sum rule cannot be decided from the present comparison.

Finally, the quadrupole columns are sparsely populated. Apart from the present
work, only the leading-order mean-field calculation in Table~\ref{tab:comp}
reports $Q_T$ as such; the isovector entry of the earlier sum-rule work is our
construction from the $\hT^{u-d}(0)$ quoted there, and differs from ours in sign,
as discussed above. None of the determinations compared gives the isoscalar
quadrupole.

One caveat on the comparison itself. Reference~\cite{Ghim:2026qnp} defines the
computed isovector mean-field quantity as $\kappa_T^{u-d}+g_T^{u-d}$ in its
Eq.~(38b), yet quotes the resulting number as $\kappa_T^{u-d}$. If the tensor
charge were not subtracted, the corresponding $\kappa_T^{u-d}$ would be smaller
by $g_T^{u-d}\simeq1$. We quote the value as published.

\subsection{Why the isoscalar tensor charge carries no leading twist}
\label{sec:isoscalar}

With the Ioffe current the two flavour channels draw on disjoint sets of
distribution amplitudes. In the chiral limit the $u$ channel is fed by $S_1$,
$S_2$ and $T_1,\dots,T_8$ alone --- every $V$ amplitude reaches it only through
the mass term of the propagator, and so carries a factor $r_u$ --- while the $d$
channel, by the selection rule of Sec.~\ref{sec:lcsr}, is fed by $V_1$, $V_2$
and $V_3$ alone. No amplitude contributes to both.

Order the contributions by the momentum-fraction integral they carry. The least
suppressed is $\mathbb{I}_1^{(2)}[1]$ of Eq.~\eqref{eq:Idef1}, whose coefficient
is
\begin{equation}
u:\;\;-2m_N\big(S_1+T_1\big)+12\,m_N T_7 ,
\qquad
d:\;\;2m_N V_1 ,
\label{eq:I2coef}
\end{equation}
the $d$ entry carrying the isospin factor that distinguishes the two
combinations. Of these, $T_1$ and $V_1$ are of twist three and $S_1$, $T_7$ of
twist four. The nucleon distribution amplitudes obey~\cite{Braun:2006hz}
\begin{equation}
2T_1(x_1,x_2,x_3)=\big[V_1-A_1\big](x_1,x_3,x_2)+\big[V_1-A_1\big](x_2,x_3,x_1),
\label{eq:isorel}
\end{equation}
and in the asymptotic limit $A_1$ vanishes while $V_1$ is symmetric, so that
$T_1=V_1$. The two twist-three terms in Eq.~\eqref{eq:I2coef} are then equal in
magnitude and opposite in sign: they cancel in the isoscalar combination
and add in the isovector one.

The consequence is that $g_T^{u+d}$ has no leading-twist contribution at the
level of $\mathbb{I}_1^{(2)}$ and is built from twist four and above, together
with whatever residue the departure of $T_1$ from $V_1$ leaves behind, whereas
$g_T^{u-d}$ receives twice the leading-twist piece. Both channels retain a
substantial higher-twist content: switching off $V_2$ and $V_3$, which reach the
$d$ channel only through $\mathbb{I}_2^{(4)}[\alpha]$, removes about two thirds
of $H_T^{d}$ at $Q^2=1\GeVsq$, so the $d$ channel is by no means saturated by
its leading-twist amplitude either.

Two independent observations support the picture. First, the isoscalar charge is
almost unchanged between the two distribution-amplitude sets, $0.39$ against
$0.38$, while the isovector charge moves from $1.27$ to $1.08$: the two sets
differ in the higher conformal-spin parameters that shape the twist-three
amplitudes, but share $f_N$, $\lambda_1$ and $\lambda_2$, so a quantity built
purely from twist four and above is expected to be the stable one. Second, and
more directly, rescaling the normalisation of the twist-three amplitudes moves
the isovector charge while leaving the isoscalar charge unchanged at the
$0.01\%$ level, which is the cancellation of Eq.~\eqref{eq:I2coef} seen in the
output rather than in the coefficients.

This also explains why the isoscalar entry is the least reliable in
Table~\ref{tab:comp}. A quantity with no leading-twist contribution is precisely
the one for which a sum rule evaluated at leading order in $\alpha_s$, with the
light-cone expansion truncated at twist six, is weakest.

\subsection{An identity in the $d$-quark channel}
\label{sec:identity}

Only the six $V$ amplitudes reach the $d$-quark channel, by the selection rule
of Sec.~\ref{sec:lcsr}, and their contributions organise themselves into the
calligraphic amplitudes of Ref.~\cite{Braun:2006hz},
$2(p\!\cdot\!x)\mathcal{V}_2=V_1-V_2-V_3$ and
$4(p\!\cdot\!x)^{2}\mathcal{V}_6=-V_1+V_2+V_3+V_4+V_5-V_6$:
\begin{align}
E_T^{d}&=16\,m_N(p\!\cdot\!x)\,\mathbb{I}_2^{(4)}[1]\,\mathcal{V}_2
 \;+\;64\,m_N\,r_d\,(p\!\cdot\!x)^{2}\,\mathbb{I}_3^{(6)}[1]\,\mathcal{V}_6 ,
\label{eq:ETdid}\\
\hT^{d}&=-16\,m_N(p\!\cdot\!x)\,\mathbb{I}_2^{(4)}[1]\,\mathcal{V}_2 ,
\label{eq:HTdid}
\end{align}
in the notation of Eqs.~\eqref{eq:Idef1} and~\eqref{eq:Idef2}, the full
expressions being Eqs.~\eqref{eq:ETd} and~\eqref{eq:HTTd} of
Appendix~\ref{app:rho}. Hence
\begin{equation}
E_T^{d}(Q^{2})+\hT^{d}(Q^{2})
 =64\,m_N\,r_d\,(p\!\cdot\!x)^{2}\,\mathbb{I}_3^{(6)}[1]\,\mathcal{V}_6 .
\label{eq:identity}
\end{equation}
The $d$-quark contributions to $E_T$ and $\hT$ are exactly equal and opposite in
the chiral limit. The breaking is triply suppressed --- by the quark mass, by
being of twist six, and by the higher power of the denominator --- and is
numerically a $4\times10^{-4}$ effect: at $Q^{2}=1\GeVsq$ we obtain
$E_T^{d}=-0.2994$ against $\hT^{d}=+0.2993$, and setting $m_d\to0$ gives
$\mp0.29931$ to every digit. It follows that
\begin{equation}
\kappa_T^{d}=-E_T^{d}+\mathcal{O}(m_d),\qquad
Q_T^{d}=-2E_T^{d}+\mathcal{O}(m_d),
\label{eq:dsector}
\end{equation}
so the $d$-quark sector carries a single independent function. The identity
holds to this accuracy on the QCD side, at every $Q^{2}$ in the working region;
what degrades it in Table~\ref{tab:2d} is the extrapolation, as quantified at
the start of this section. Because Eq.~\eqref{eq:identity} isolates
$\mathcal{V}_6$ multiplied by the quark mass, it also offers a direct window on
the size of the twist-six contribution in the tensor channel.

\subsection{Test of the large-$N_c$ relation}
\label{sec:largeNc}

In the mean-field picture $E_T^{u-d}$ and $\hT^{u-d}$ are both
$\mathcal{O}(N_c^{3})$ while their combination is $\mathcal{O}(N_c)$, so the
cancellation $2\hT^{u-d}=-E_T^{u-d}$ is enforced to two powers of
$N_c$~\cite{Ghim:2025gqo}. Light-cone QCD sum rules make no assumption about $N_c$,
so our results provide an independent test. Using Eq.~\eqref{eq:identity} the
relation reduces to $2\hT^{u}=-E_T^{u}-E_T^{d}$, so any deviation resides
entirely in the $u$ channel.

We find
\begin{equation}
\frac{2\hT^{u-d}}{-E_T^{u-d}}=0.64(16)\quad\text{(Set~I)},\qquad
0.67(15)\quad\text{(Set~II)},
\label{eq:largeNc}
\end{equation}
against the value of unity that the relation predicts. The sign and the order of
magnitude come out right, with the central values lying about a third below
unity, and the deviation is $2.2\sigma$ in both sets. The two sets agree with
each other to within three per cent, so whatever is responsible is systematic
rather than a fit artefact; but at $2.2\sigma$, and with the quoted
uncertainties covering the $s_0$ spread alone, this is a mild tension rather
than a violation. It is worth reading together with the fact that a sum rule
evaluated at leading order in $\alpha_s$ and truncated at twist six is itself
least reliable for combinations in which a cancellation occurs, which is exactly
the situation here.

Expressed through the $u$ channel with the help of Eq.~\eqref{eq:identity},
the relation requires
$\hT^{u}=-1.86(56)$ and $-1.79(48)$ for the two sets, against the $-0.50(25)$
and $-0.66(20)$ we obtain, a $2.2\sigma$ difference in both cases; the $d$
channel satisfies its part exactly, by construction. Those two values are the
isospin reconstruction of Table~\ref{tab:ff}; fixing $\hT^{d}$ from
Eq.~\eqref{eq:identity} instead would shift them to $-0.64$ and $-0.73$, well
inside the quoted uncertainty, so the conclusion does not depend on which route
is taken. Whatever the status of the relation, the resulting flavour asymmetry
is pronounced and comfortably survives the uncertainties: $|\hT/E_T|$ is unity
for the $d$ quark, exactly, and $0.09(5)$ and $0.13(5)$ for the $u$ quark in the
two sets, more than fifteen standard deviations away from unity. This is the
direct reason the tensor quadrupole moment sits almost entirely on the $d$
quark, as Table~\ref{tab:2d} shows.

\subsection{Impact-parameter shapes}
\label{sec:shapes}

The multipole moments of Table~\ref{tab:2d} have a direct interpretation in the
impact-parameter plane. Fourier transforming the matrix element of the tensor
current with respect to the transverse momentum transfer,
\begin{equation}
\rho_q(\boldsymbol{b}_\perp)=\int\!\frac{d^2\Delta_\perp}{(2\pi)^2}\,
e^{-i\boldsymbol{\Delta}_\perp\cdot\boldsymbol{b}_\perp}\,
\frac{1}{2P^+}\,\mathcal{M}_{\mathrm{FFs}}[i\sigma^{+j}] ,
\label{eq:FT}
\end{equation}
gives the transverse distribution of quarks of a definite transverse
polarisation $\boldsymbol{s}_\perp$ in a nucleon of transverse polarisation
$\boldsymbol{S}_\perp$. Here $v^{\pm}=(v^{0}\pm v^{3})/\sqrt2$ are light-cone
components, $j=1,2$ is a transverse index, $P=(p'+p)/2$ so that $P^{+}$ is the
average light-cone momentum, $\boldsymbol{\Delta}_\perp$ is the transverse
momentum transfer with $t=-|\boldsymbol{\Delta}_\perp|^{2}$, and
$\mathcal{M}_{\mathrm{FFs}}[i\sigma^{+j}]$ denotes the matrix element of
Eq.~\eqref{eq:param} with the tensor indices taken as $\mu\nu=+j$. Both
polarisations are needed. In the two-dimensional
multipole expansion of Ref.~\cite{Kim:2024ibz} the monopole and the quadrupole
sit on nucleon spin operators and the dipole on the unit operator, so in an
unpolarised nucleon only the dipole survives; that limit is the Burkardt
distortion~\cite{Burkardt:2005hp} and it carries $\kappa_T$ alone. To display
all three moments we therefore take the nucleon polarised as well, along the
same axis as the quarks.

Contracting the transverse index $j$ of the operator with the quark
polarisation and taking $\boldsymbol{s}_\perp=\boldsymbol{S}_\perp=\hat{x}$, the
three structures of Ref.~\cite{Kim:2024ibz} reduce to an isotropic term, a term
odd in $b_y$, and a $\cos2\phi_b$ term, so that
\begin{equation}
\rho_q(b_\perp,\phi_b)=\rho_0(b_\perp)
\Big[1+a_1(b_\perp)\sin\phi_b+a_2(b_\perp)\cos2\phi_b\Big],
\label{eq:rho_impact}
\end{equation}
with $\phi_b$ measured from the polarisation axis. No higher harmonics occur,
since the expansion stops at the quadrupole (Sec.~\ref{sec:multipole}). The
radial profiles $\rho_0$, $a_1$ and $a_2$ require the form factors at all $Q^2$
and are not determined here. Two moments of the distribution are determined,
however, and they depend on the profiles only through integrals. With
$\langle O\rangle_q=\int d^2b_\perp\,O\,\rho_q\big/\!\int d^2b_\perp\,\rho_q$,
the angular integrations in Eq.~\eqref{eq:rho_impact} give
\begin{equation}
\langle b_y\rangle_q=\frac{\int_0^\infty db_\perp\,b_\perp^{2}\,\rho_0\,a_1}
{2\int_0^\infty db_\perp\,b_\perp\,\rho_0},\qquad
\big\langle b_x^{2}-b_y^{2}\big\rangle_q=\frac{\int_0^\infty db_\perp\,b_\perp^{3}\,\rho_0\,a_2}
{2\int_0^\infty db_\perp\,b_\perp\,\rho_0},
\label{eq:moments_profiles}
\end{equation}
so that the displacement is fixed by $a_1$ alone and the deformation by $a_2$
alone. The same moments follow directly from Eq.~\eqref{eq:FT}: weighting
$\rho_q$ with $b_y$ or with $b_x^2-b_y^2$ and integrating over the plane amounts
to differentiating the matrix element once or twice with respect to
$\boldsymbol{\Delta}_\perp$ at $\boldsymbol{\Delta}_\perp=0$, where only the form
factors at $t=0$ survive. The normalisation is set by the monopole, the first
derivative picks out the dipole term and the traceless part of the second the
quadrupole term, so that
\begin{equation}
\langle b_y\rangle_q=\frac{\kappa_T^{q}}{2m_N\,g_T^{q}},
\qquad
\big\langle b_x^{2}-b_y^{2}\big\rangle_q=\frac{Q_T^{q}}{2m_N^{2}\,g_T^{q}} ,
\label{eq:shape_moments}
\end{equation}
independently of the profiles.
The first is the displacement of the centroid perpendicular to the polarisation
axis, the light-front analogue of the shift generated by the anomalous magnetic
moment in the vector channel; the second is the quadrupole deformation, positive
when the distribution is elongated along the polarisation axis and negative when
it is elongated across it. The monopole $g_T$ sets the normalisation and appears
in both denominators, so it is the relative, not the absolute, size of
$\kappa_T$ and $Q_T$ that shapes the distribution.

\begin{figure}[htb]
\centering
\includegraphics[width=\textwidth]{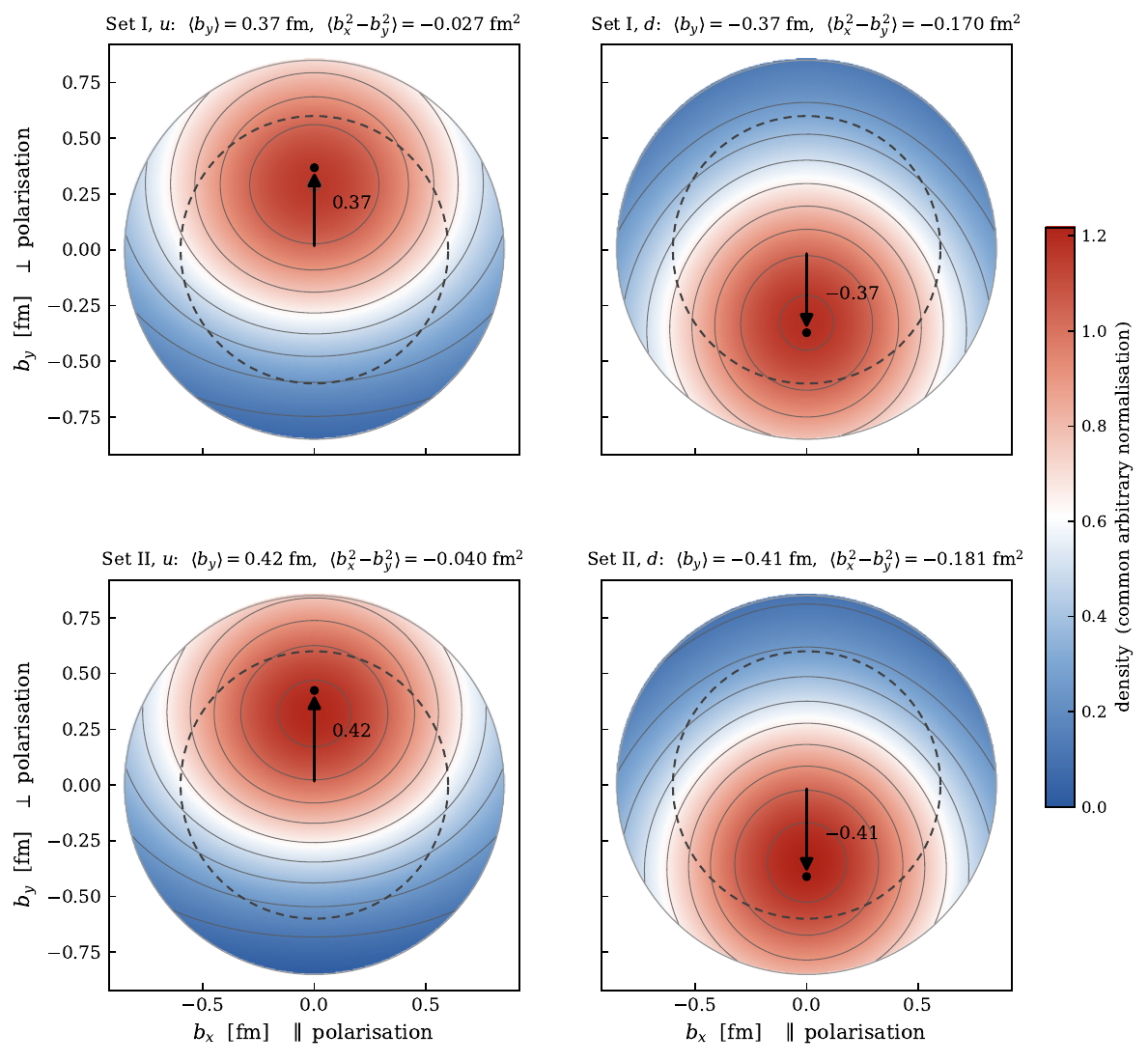}
\caption{Impact-parameter distributions of $u$ (left) and $d$ (right) quarks
polarised along $+x$ in a nucleon polarised along the same direction, from
Eq.~\eqref{eq:rho_impact} with the moments of Eq.~\eqref{eq:shape_params}, a
Gaussian profile of width $R=0.60$~fm and the profiles $a_1$ and $a_2$ described
in the text. Top row: Set~I; bottom row: Set~II. The
colour map shows the density on a common scale, the grey curves are isodensity
contours, the dashed circle is the undisplaced, undeformed reference and the
arrow marks $\langle b_y\rangle$ in fm. The two flavours are displaced in
opposite directions by nearly equal amounts, while both distributions are
elongated across the polarisation axis, the $d$ quark some five to six times
more strongly. The overall normalisation of each panel is set by $g_T^{q}$,
which is negative for the $d$ quark, and is not displayed; only the shapes are
compared. The density is shown within $|b_\perp|\le0.85$~fm. Beyond
$|b_\perp|\simeq0.9$~fm the linearly and quadratically rising profiles $a_1$ and
$a_2$, which are a display choice, begin to outweigh the isotropic term, and the
displayed density would no longer be positive.}
\label{fig:quark_shapes}
\end{figure}

With the moments of Table~\ref{tab:2d} we obtain
\begin{equation}
\begin{aligned}
\text{Set~I}:&\quad \langle b_y\rangle_{u,d}=0.37,\,-0.37\ \text{fm}, &\quad
\big\langle b_x^{2}-b_y^{2}\big\rangle_{u,d}&=-0.027,\,-0.170\ \text{fm}^{2},\\[3pt]
\text{Set~II}:&\quad \langle b_y\rangle_{u,d}=0.42,\,-0.41\ \text{fm}, &\quad
\big\langle b_x^{2}-b_y^{2}\big\rangle_{u,d}&=-0.040,\,-0.181\ \text{fm}^{2} .
\end{aligned}
\label{eq:shape_params}
\end{equation}
Figure~\ref{fig:quark_shapes} shows the corresponding distributions. For display
we take a Gaussian isotropic profile, $\rho_0\propto\exp(-b_\perp^2/2R^2)$, and
the lowest profiles compatible with regularity at the origin,
$a_1=c_1b_\perp$ and $a_2=c_2b_\perp^2$: for the density to be smooth at
$b_\perp=0$, the $\sin\phi_b$ and $\cos2\phi_b$ terms must vanish there as
$b_\perp$ and $b_\perp^2$, respectively. Equation~\eqref{eq:moments_profiles}
then fixes $c_1=\langle b_y\rangle/R^2$ and
$c_2=\langle b_x^2-b_y^2\rangle/(4R^4)$, so that the displayed distributions
reproduce the moments of Eq.~\eqref{eq:shape_params} exactly, whatever the value
of $R$. We take $R=0.60$~fm, of the order of the transverse size of the nucleon.
The width affects the appearance of the figure, but neither the moments nor the
conclusions drawn from them below, which rest on Eq.~\eqref{eq:shape_moments}
alone.

The two flavours are displaced in opposite directions by nearly equal amounts.
This follows from the ratios $\kappa_T/g_T$, which come out $3.5$ and $4.0$ for
the $u$ quark in the two sets and $-3.5$ and $-3.9$ for the $d$ quark: $g_T^{d}$
is negative while $\kappa_T^{d}$ is positive, and it is that sign mismatch,
rather than any difference in the magnitude of the distortion, that reverses the
shift. The deformations, in contrast, have the same sign for both flavours, both
distributions being elongated across the polarisation axis, but they differ by a
factor of five to six: $Q_T/g_T$ is $-1.2$ and $-1.8$ for the $u$ quark against
$-7.7$ and $-8.2$ for the $d$ quark. This is the impact-parameter form of the
flavour hierarchy of Sec.~\ref{sec:largeNc}, where $|\hT/E_T|$ is unity for the
$d$ quark and an order of magnitude smaller for the $u$ quark. The two
distribution-amplitude sets give the same picture throughout.

\section{Summary}
\label{sec:summary}

The tensor current is odd under charge conjugation, so the quantities it
defines measure the difference between quark and antiquark contributions, and
its three chiral-odd form factors organise themselves into a monopole, a dipole
and a quadrupole. We have computed all three for the nucleon within light-cone
QCD sum rules, resolved by flavour, over $1\le Q^2\le10\GeVsq$, and assembled
the corresponding multipole moments from their values at $Q^2=0$. The flavour
components of the quadrupole moment and the isoscalar channel had not previously
been obtained in this framework.

Three results stand apart from the numbers themselves. The first is
methodological. The Lorentz structures generated by the operator product are not
independent until the Dirac strings are brought to a canonical order. Eight
independent antisymmetric structures survive the reduction; one of them does not
appear on the hadronic side at all, one mixes two form factors, and of the six
that are clean, four return the tensor charge, so that exactly three suffice to
give the three form factors separately. The structure from which the tensor
charge has usually been extracted is not among the eight. Before ordering the
structures are linearly related, so the coefficient assigned to any one of them
is not fixed by the content of the expression alone; the basis used here is
therefore different from that of earlier sum-rule work, independently of the
interpolating current or the fit range.

The second is an identity. Only the vector distribution amplitudes reach the
$d$-quark channel with the Ioffe current, and their contributions to $E_T$ and
$\hT$ are equal and opposite in the chiral limit, the breaking being
proportional to the quark mass and to the twist-six amplitude $\mathcal{V}_6$.
Numerically the breaking is a $4\times10^{-4}$ effect, and setting the quark
mass to zero reproduces the cancellation to every digit. The $d$-quark sector
therefore carries a single independent function, with $\kappa_T^{d}=-E_T^{d}$
and $Q_T^{d}=-2E_T^{d}$ on the QCD side, up to the quark mass. The extraction
confirms this: fitting $E_T+2\hT$ as a function of $Q^{2}$ returns
$\kappa_T^{d}=-E_T^{d}$ to four digits in both distribution-amplitude sets. The
$9\%$ and $5\%$ by which the quoted $Q_T^{d}$ departs from $2\kappa_T^{d}$ is
not a failure of the identity but the cost of reaching $\hT^{d}$ through the two
isospin combinations, and it sets the extraction systematic on the $\hT$ and
$Q_T$ entries.

The third concerns the isoscalar channel. The two flavour channels draw on
disjoint sets of distribution amplitudes, and their leading-twist contributions
to the tensor charge are equal and opposite, so that they cancel in the
isoscalar combination and add in the isovector one. The isoscalar tensor charge
thus has no leading-twist contribution at the level of the least suppressed
integral. That explains both why it comes out small and why it is the least
reliable entry in the comparison: a quantity built essentially from twist four
and above is precisely the one for which a sum rule
evaluated at leading order in $\alpha_s$, with the light-cone expansion
truncated at twist six, is weakest. Two independent checks support the
mechanism, the stability of the isoscalar charge against the choice of
distribution-amplitude set and its insensitivity to the twist-three
normalisation. 
None of these three would have been visible without resolving the flavours: each
is a statement about one channel, or about the relation between the two.

The same is true of the picture the moments produce. Carried to the
impact-parameter plane, and taken with the quark and the nucleon polarised along
the same transverse axis, they displace the two flavour distributions in
opposite directions by nearly equal amounts, $\langle b_y\rangle=0.37$ and
$-0.37$~fm, and elongate both across that axis, the $d$ quark some five to six
times more strongly. The opposite displacements do not come from a difference in
the distortion itself but from the sign of $g_T$, which is negative for the $d$
quark while $\kappa_T$ is positive for both; and the disparity in the
deformations is the impact-parameter form of the flavour asymmetry in
$|\hT/E_T|$, which is unity for the $d$ quark and an order of magnitude smaller
for the $u$ quark. Both distribution-amplitude sets give the same picture.

Against the other determinations the isovector tensor charge sits inside the
existing band and the isovector dipole moment in the middle of it, while the
isoscalar charge lies below for the reason just given and the isoscalar dipole
falls just under the two soliton values. The dipole moment shows the widest
spread of any quantity in the comparison, which is itself a consequence of its
being the residual of a cancellation enforced to two powers of $N_c$; it has
been computed once on the lattice and has no phenomenological determination at
all. Among the determinations compared, only the mean-field calculation gives
the quadrupole moment as such, and none gives the isoscalar quadrupole, so this
is the entry that would benefit most from an independent calculation. Tested
against our results, the mean-field relation $2\hT^{u-d}=-E_T^{u-d}$ holds in
sign and in order of magnitude, the ratio coming out $0.64(16)$ and $0.67(15)$
against the predicted unity, a $2.2\sigma$ deviation that resides entirely in
the $u$ channel.

\clearpage
\appendix

\section{QCD-side results}
\label{app:rho}

The invariant functions multiplying the independent structures of
Eq.~\eqref{eq:projection}, before the Borel transformation and the continuum
subtraction, are collected here. It is convenient to abbreviate the
momentum-fraction integrals as
\begin{align}
\mathbb{I}_1^{(n)}[w]&=m_N^{\,n-2}\!\int_0^1\!\!d\xi\;\frac{w}{(q-\xi p)^{n}},
\label{eq:Idef1}\\
\mathbb{I}_2^{(n)}[w]&=m_N^{\,n-2}\!\int_0^1\!\!d\xi\int_0^{\xi}\!\!d\alpha\;
  \frac{w}{(q-\alpha p)^{n}},\\
\mathbb{I}_3^{(n)}[w]&=m_N^{\,n-2}\!\int_0^1\!\!d\xi\int_0^{\xi}\!\!d\alpha
  \int_0^{\alpha}\!\!d\beta\;\frac{w}{(q-\beta p)^{n}} ,
\label{eq:Idef2}
\end{align}
the weight $w$ being $1$, $\alpha$, $\beta$ or $\beta^{2}$. The factor
$m_N^{\,n-2}$ is included so that all three carry the same mass dimension; with
this normalisation, and writing $r_q=m_q/m_N$, every term below is an integer
multiple of $m_N$ times one of them. The distribution amplitudes are those of
Refs.~\cite{Braun:2000kw,Braun:2006hz}, and the quark masses enter only through
the mass term of the light-quark propagator, Eq.~\eqref{eq:prop}. The
$u$-channel amplitudes are evaluated at $x_2$ and the $d$-channel ones at $x_3$,
as required by the slot structure of Sec.~\ref{sec:lcsr}; the arguments are
suppressed below. The isoscalar and isovector combinations are obtained by
adding the two channels with a relative sign $+1$ and $-1$ respectively.

\subsection{Tensor form factor \texorpdfstring{$H_T$}{H\_T}}

\begin{align}
\label{eq:HTu}
\Pi^{u}_{H_T}=2m_N\Big\{\;
 &-\Big[\mathbb{I}_{1}^{(2)}[1]+\mathbb{I}_{2}^{(4)}[\alpha]\Big]S_{1}
  +\mathbb{I}_{2}^{(4)}[\alpha]\,S_{2}\nonumber\\[2pt]
 &-\Big[\mathbb{I}_{1}^{(2)}[1]+4\,\mathbb{I}_{2}^{(4)}[\alpha]
        +3\,\mathbb{I}_{3}^{(4)}[1]\Big]T_{1}
  +\mathbb{I}_{1}^{(4)}[1]\,T_{1}^{M}\nonumber\\[2pt]
 &+\Big[3\,\mathbb{I}_{2}^{(4)}[\alpha]+4\,\mathbb{I}_{3}^{(6)}[\beta^{2}]
        +4\,\mathbb{I}_{3}^{(4)}[1]\Big]\big(T_{2}+T_{5}\big)\nonumber\\[2pt]
 &-\Big[2\,\mathbb{I}_{2}^{(4)}[\alpha]+4\,\mathbb{I}_{3}^{(6)}[\beta^{2}]
        +\mathbb{I}_{3}^{(4)}[1]\Big]T_{3}
  -\Big[4\,\mathbb{I}_{3}^{(6)}[\beta^{2}]+\mathbb{I}_{3}^{(4)}[1]\Big]T_{4}
  \nonumber\\[2pt]
 &-3\,\mathbb{I}_{3}^{(4)}[1]\,T_{6}
  +\Big[6\,\mathbb{I}_{1}^{(2)}[1]+10\,\mathbb{I}_{2}^{(4)}[\alpha]
        +4\,\mathbb{I}_{3}^{(6)}[\beta^{2}]+7\,\mathbb{I}_{3}^{(4)}[1]\Big]T_{7}+\Big[4\,\mathbb{I}_{3}^{(6)}[\beta^{2}]+7\,\mathbb{I}_{3}^{(4)}[1]\Big]T_{8}
  \nonumber\\[4pt]
 &+r_u\Big(-\Big[\mathbb{I}_{2}^{(4)}[1]+2\,\mathbb{I}_{3}^{(6)}[\beta]\Big]V_{1}
   +2\,\mathbb{I}_{3}^{(6)}[\beta]\big(V_{2}+V_{3}-V_{6}\big)+\Big[\mathbb{I}_{2}^{(4)}[1]+2\,\mathbb{I}_{3}^{(6)}[\beta]\Big]
   \big(V_{4}+V_{5}\big)\Big)\Big\},
\\[4pt]
\Pi^{d}_{H_T}=2m_N\Big\{\;
 &\Big[\mathbb{I}_{1}^{(2)}[1]+\mathbb{I}_{2}^{(4)}[\alpha]\Big]V_{1}
  -\mathbb{I}_{2}^{(4)}[\alpha]\big(V_{2}+V_{3}\big)
  -\mathbb{I}_{1}^{(4)}[1]\,V_{1}^{M}\nonumber\\[4pt]
 &+r_d\Big(\Big[\mathbb{I}_{2}^{(4)}[1]+2\,\mathbb{I}_{3}^{(6)}[\beta]\Big]
   \big(V_{1}-V_{4}-V_{5}\big)-2\,\mathbb{I}_{3}^{(6)}[\beta]\big(V_{2}+V_{3}-V_{6}\big)\Big)\Big\}.
\label{eq:HTd}
\end{align}

Two features of Eqs.~\eqref{eq:HTu} and~\eqref{eq:HTd} are used in
Sec.~\ref{sec:isoscalar}. The $u$ channel receives the $V$ amplitudes only
through $r_u$, so in the chiral limit it is fed by $S$ and $T$ alone, while the
$d$ channel is fed by $V$ alone. And the coefficients of the least suppressed
integral $\mathbb{I}_1^{(2)}$ are $-\big(S_1+T_1\big)+6\,T_7$ in the $u$ channel
against $V_1$ in the $d$ channel.

\subsection{Tensor form factor \texorpdfstring{$E_T$}{E\_T}}

\begin{align}
\label{eq:ETu}
\Pi^{u}_{E_T}=8m_N\Big\{\;
 &\mathbb{I}_{2}^{(4)}[1]\Big(S_{1}-S_{2}+2\,T_{1}+T_{2}-3\,T_{5}-4\,T_{7}\Big)
   +4\,\mathbb{I}_{3}^{(6)}[\beta]
   \Big(-T_{2}+T_{3}+T_{4}-T_{5}-T_{7}-T_{8}\Big)\nonumber\\[4pt]
 &+2\,r_u\,\mathbb{I}_{3}^{(6)}[1]
   \Big(V_{1}-V_{2}-V_{3}-V_{4}-V_{5}+V_{6}\Big)\Big\},
\\[4pt]
\Pi^{d}_{E_T}=8m_N\Big\{\;
 &\mathbb{I}_{2}^{(4)}[1]\Big(V_{1}-V_{2}-V_{3}\Big)-2\,r_d\,\mathbb{I}_{3}^{(6)}[1]
   \Big(V_{1}-V_{2}-V_{3}-V_{4}-V_{5}+V_{6}\Big)\Big\}.
\label{eq:ETd}
\end{align}

\subsection{Tensor form factor \texorpdfstring{$\hT$}{Htilde\_T}}

\begin{align}
\Pi^{u}_{\hT}&=16\,m_N\,\mathbb{I}_{2}^{(4)}[1]
  \Big(-T_{1}+T_{3}+T_{7}\Big),
\label{eq:HTTu}\\[4pt]
\Pi^{d}_{\hT}&=-8\,m_N\,\mathbb{I}_{2}^{(4)}[1]
  \Big(V_{1}-V_{2}-V_{3}\Big).
\label{eq:HTTd}
\end{align}

The identity of Sec.~\ref{sec:identity} can now be read off directly. Both
$\hT^{d}$ and the mass-independent part of $E_T^{d}$ are proportional to
$V_1-V_2-V_3=2(p\!\cdot\!x)\mathcal{V}_2$ with the same prefactor and opposite
sign, so they cancel term by term. What survives is the mass term of
$E_T^{d}$, whose amplitude combination is
$V_1-V_2-V_3-V_4-V_5+V_6=-4(p\!\cdot\!x)^{2}\mathcal{V}_6$, giving
Eq.~\eqref{eq:identity}.

The same combination appears in $E_T^{u}$, with the opposite overall sign and
with $r_u$ in place of $r_d$. In the isospin limit the two therefore cancel in
the isoscalar combination and add in the isovector one, although the effect is
numerically negligible, being both mass-suppressed and of twist six.

\bibliographystyle{elsarticle-num}
\bibliography{Tensor_multipole_moments_of_the_nucleon.bib}

@article{Jaffe:1991kp,
    author = "Jaffe, R. L. and Ji, Xiang-Dong",
    title = "{Chiral odd parton distributions and polarized Drell-Yan}",
    reportNumber = "MIT-CTP-1952",
    doi = "10.1103/PhysRevLett.67.552",
    journal = "Phys. Rev. Lett.",
    volume = "67",
    pages = "552--555",
    year = "1991"
}

@article{Jaffe:1991ra,
    author = "Jaffe, R. L. and Ji, Xiang-Dong",
    title = "{Chiral odd parton distributions and Drell-Yan processes}",
    reportNumber = "MIT-CTP-2005",
    doi = "10.1016/0550-3213(92)90110-W",
    journal = "Nucl. Phys. B",
    volume = "375",
    pages = "527--560",
    year = "1992"
}

@article{Ralston:1979ys,
    author = "Ralston, John P. and Soper, Davison E.",
    title = "{Production of Dimuons from High-Energy Polarized Proton Proton Collisions}",
    reportNumber = "OITS-100",
    doi = "10.1016/0550-3213(79)90082-8",
    journal = "Nucl. Phys. B",
    volume = "152",
    pages = "109",
    year = "1979"
}

@article{Barone:2001sp,
    author = "Barone, Vincenzo and Drago, Alessandro and Ratcliffe, Philip G.",
    title = "{Transverse polarisation of quarks in hadrons}",
    eprint = "hep-ph/0104283",
    archivePrefix = "arXiv",
    doi = "10.1016/S0370-1573(01)00051-5",
    journal = "Phys. Rept.",
    volume = "359",
    pages = "1--168",
    year = "2002"
}

@article{Barone:2010zz,
    author = "Barone, Vincenzo and Bradamante, Franco and Martin, Anna",
    title = "{Transverse-spin and transverse-momentum effects in high-energy processes}",
    eprint = "1011.0909",
    archivePrefix = "arXiv",
    primaryClass = "hep-ph",
    doi = "10.1016/j.ppnp.2010.07.003",
    journal = "Prog. Part. Nucl. Phys.",
    volume = "65",
    pages = "267--333",
    year = "2010"
}

@article{Soffer:1994ww,
    author = "Soffer, Jacques",
    title = "{Positivity constraints for spin dependent parton distributions}",
    eprint = "hep-ph/9409254",
    archivePrefix = "arXiv",
    reportNumber = "CPT-94-P-3059",
    doi = "10.1103/PhysRevLett.74.1292",
    journal = "Phys. Rev. Lett.",
    volume = "74",
    pages = "1292--1294",
    year = "1995"
}

@article{Pospelov:2005pr,
    author = "Pospelov, Maxim and Ritz, Adam",
    title = "{Electric dipole moments as probes of new physics}",
    eprint = "hep-ph/0504231",
    archivePrefix = "arXiv",
    doi = "10.1016/j.aop.2005.04.002",
    journal = "Annals Phys.",
    volume = "318",
    pages = "119--169",
    year = "2005"
}

@article{Bhattacharya:2015esa,
    author = "Bhattacharya, Tanmoy and Cirigliano, Vincenzo and Gupta, Rajan and Lin, Huey-Wen and Yoon, Boram",
    title = "{Neutron Electric Dipole Moment and Tensor Charges from Lattice QCD}",
    eprint = "1506.04196",
    archivePrefix = "arXiv",
    primaryClass = "hep-lat",
    reportNumber = "LA-UR-15-24210",
    doi = "10.1103/PhysRevLett.115.212002",
    journal = "Phys. Rev. Lett.",
    volume = "115",
    number = "21",
    pages = "212002",
    year = "2015"
}

@article{Pitschmann:2014jxa,
    author = "Pitschmann, Mario and Seng, Chien-Yeah and Roberts, Craig D. and Schmidt, Sebastian M.",
    title = "{Nucleon tensor charges and electric dipole moments}",
    eprint = "1411.2052",
    archivePrefix = "arXiv",
    primaryClass = "nucl-th",
    reportNumber = "ACFI-T14-23",
    doi = "10.1103/PhysRevD.91.074004",
    journal = "Phys. Rev. D",
    volume = "91",
    pages = "074004",
    year = "2015"
}

@article{Liu:2017olr,
    author = "Liu, Tianbo and Zhao, Zhiwen and Gao, Haiyan",
    title = "{Experimental constraint on quark electric dipole moments}",
    eprint = "1704.00113",
    archivePrefix = "arXiv",
    primaryClass = "hep-ph",
    doi = "10.1103/PhysRevD.97.074018",
    journal = "Phys. Rev. D",
    volume = "97",
    number = "7",
    pages = "074018",
    year = "2018"
}

@article{Courtoy:2015haa,
    author = "Courtoy, Aurore and Bae{\ss}ler, Stefan and Gonz{\'a}lez-Alonso, Mart{\'\i}n and Liuti, Simonetta",
    title = "{Beyond-Standard-Model Tensor Interaction and Hadron Phenomenology}",
    eprint = "1503.06814",
    archivePrefix = "arXiv",
    primaryClass = "hep-ph",
    doi = "10.1103/PhysRevLett.115.162001",
    journal = "Phys. Rev. Lett.",
    volume = "115",
    pages = "162001",
    year = "2015"
}

@article{Bhattacharya:2011qm,
    author = "Bhattacharya, Tanmoy and Cirigliano, Vincenzo and Cohen, Saul D. and Filipuzzi, Alberto and Gonzalez-Alonso, Martin and Graesser, Michael L. and Gupta, Rajan and Lin, Huey-Wen",
    title = "{Probing Novel Scalar and Tensor Interactions from (Ultra)Cold Neutrons to the LHC}",
    eprint = "1110.6448",
    archivePrefix = "arXiv",
    primaryClass = "hep-ph",
    reportNumber = "LA-UR-11-11460, NT@UW-11-16, IFIC-11-57, FTUV-11-1007, NPAC-11-14",
    doi = "10.1103/PhysRevD.85.054512",
    journal = "Phys. Rev. D",
    volume = "85",
    pages = "054512",
    year = "2012"
}

@article{Airapetian:2010ds,
    author = "Airapetian, A. and others",
    collaboration = "HERMES",
    title = "{Effects of transversity in deep-inelastic scattering by polarized protons}",
    eprint = "1006.4221",
    archivePrefix = "arXiv",
    primaryClass = "hep-ex",
    reportNumber = "DESY-10-087",
    doi = "10.1016/j.physletb.2010.08.012",
    journal = "Phys. Lett. B",
    volume = "693",
    pages = "11--16",
    year = "2010"
}

@article{Adolph:2012sn,
    author = "Adolph, C. and others",
    collaboration = "COMPASS",
    title = "{Experimental investigation of transverse spin asymmetries in muon-p SIDIS processes: Collins asymmetries}",
    eprint = "1205.5121",
    archivePrefix = "arXiv",
    primaryClass = "hep-ex",
    reportNumber = "CERN-PH-EP-2012-146",
    doi = "10.1016/j.physletb.2012.09.055",
    journal = "Phys. Lett. B",
    volume = "717",
    pages = "376--382",
    year = "2012"
}

@article{Vossen:2011fk,
    author = "Vossen, A. and others",
    collaboration = "Belle",
    title = "{Observation of transverse polarization asymmetries of charged pion pairs in $e^+e^-$ annihilation near $\sqrt{s}=10.58$ GeV}",
    eprint = "1104.2425",
    archivePrefix = "arXiv",
    primaryClass = "hep-ex",
    reportNumber = "BELLE-DRAFT-2011-5, KEK-PREPRINT-2010-53",
    doi = "10.1103/PhysRevLett.107.072004",
    journal = "Phys. Rev. Lett.",
    volume = "107",
    pages = "072004",
    year = "2011"
}

@article{Bacchetta:2012ty,
    author = "Bacchetta, Alessandro and Courtoy, A. and Radici, Marco",
    title = "{First extraction of valence transversities in a collinear framework}",
    eprint = "1212.3568",
    archivePrefix = "arXiv",
    primaryClass = "hep-ph",
    doi = "10.1007/JHEP03(2013)119",
    journal = "JHEP",
    volume = "03",
    pages = "119",
    year = "2013"
}

@article{Radici:2015mwa,
    author = "Radici, Marco and Courtoy, A. and Bacchetta, Alessandro and Guagnelli, Marco",
    title = "{Improved extraction of valence transversity distributions from inclusive dihadron production}",
    eprint = "1503.03495",
    archivePrefix = "arXiv",
    primaryClass = "hep-ph",
    doi = "10.1007/JHEP05(2015)123",
    journal = "JHEP",
    volume = "05",
    pages = "123",
    year = "2015"
}

@article{Anselmino:2013vqa,
    author = "Anselmino, M. and Boglione, M. and D'Alesio, U. and Melis, S. and Murgia, F. and Prokudin, A.",
    title = "{Simultaneous extraction of transversity and Collins functions from new SIDIS and e+e- data}",
    eprint = "1303.3822",
    archivePrefix = "arXiv",
    primaryClass = "hep-ph",
    reportNumber = "JLAB-THY-13-1704",
    doi = "10.1103/PhysRevD.87.094019",
    journal = "Phys. Rev. D",
    volume = "87",
    pages = "094019",
    year = "2013"
}

@article{Kang:2015msa,
    author = "Kang, Zhong-Bo and Prokudin, Alexei and Sun, Peng and Yuan, Feng",
    title = "{Extraction of Quark Transversity Distribution and Collins Fragmentation Functions with QCD Evolution}",
    eprint = "1505.05589",
    archivePrefix = "arXiv",
    primaryClass = "hep-ph",
    reportNumber = "JLAB-THY-15-2044",
    doi = "10.1103/PhysRevD.93.014009",
    journal = "Phys. Rev. D",
    volume = "93",
    number = "1",
    pages = "014009",
    year = "2016"
}

@article{Lin:2017stx,
    author = "Lin, Huey-Wen and Melnitchouk, W. and Prokudin, Alexei and Sato, N. and Shows, H.",
    title = "{First Monte Carlo Global Analysis of Nucleon Transversity with Lattice QCD Constraints}",
    eprint = "1710.09858",
    archivePrefix = "arXiv",
    primaryClass = "hep-ph",
    reportNumber = "MSUHEP-17-016, JLAB-THY-17-2574",
    doi = "10.1103/PhysRevLett.120.152502",
    journal = "Phys. Rev. Lett.",
    volume = "120",
    number = "15",
    pages = "152502",
    year = "2018"
}

@article{Cocuzza:2023oam,
    author = "Cocuzza, C. and Metz, A. and Pitonyak, D. and Prokudin, A. and Sato, N. and Seidl, R.",
    collaboration = "JAM",
    title = "{Transversity Distributions and Tensor Charges of the Nucleon: Extraction from Dihadron Production and Their Universal Nature}",
    eprint = "2306.12998",
    archivePrefix = "arXiv",
    primaryClass = "hep-ph",
    reportNumber = "JLAB-THY-23-3859",
    doi = "10.1103/PhysRevLett.132.091901",
    journal = "Phys. Rev. Lett.",
    volume = "132",
    number = "9",
    pages = "091901",
    year = "2024"
}

@article{Accardi:2012qut,
    author = "Accardi, A. and others",
    editor = "Deshpande, A. and Meziani, Z. E. and Qiu, J. W.",
    title = "{Electron Ion Collider: The Next QCD Frontier}: {Understanding the glue that binds us all}",
    eprint = "1212.1701",
    archivePrefix = "arXiv",
    primaryClass = "nucl-ex",
    reportNumber = "BNL-98815-2012-JA, JLAB-PHY-12-1652",
    doi = "10.1140/epja/i2016-16268-9",
    journal = "Eur. Phys. J. A",
    volume = "52",
    number = "9",
    pages = "268",
    year = "2016"
}

@article{AbdulKhalek:2021gbh,
    author = "Abdul Khalek, R. and others",
    title = "{Science Requirements and Detector Concepts for the Electron-Ion Collider}: {EIC Yellow Report}",
    eprint = "2103.05419",
    archivePrefix = "arXiv",
    primaryClass = "physics.ins-det",
    reportNumber = "BNL-220990-2021-FORE, JLAB-PHY-21-3198, LA-UR-21-20953",
    doi = "10.1016/j.nuclphysa.2022.122447",
    journal = "Nucl. Phys. A",
    volume = "1026",
    pages = "122447",
    year = "2022"
}

@article{Gockeler:2005cj,
    author = "Gockeler, M. and Hagler, Ph. and Horsley, R. and Pleiter, D. and Rakow, Paul E. L. and Schafer, A. and Schierholz, G. and Zanotti, J. M.",
    collaboration = "QCDSF, UKQCD",
    title = "{Quark helicity flip generalized parton distributions from two-flavor lattice QCD}",
    eprint = "hep-lat/0507001",
    archivePrefix = "arXiv",
    reportNumber = "DESY-05-107, EDINBURGH-2005-05, MPP-2005-60",
    doi = "10.1016/j.physletb.2005.09.002",
    journal = "Phys. Lett. B",
    volume = "627",
    pages = "113--123",
    year = "2005"
}

@article{Bhattacharya:2015wna,
    author = "Bhattacharya, Tanmoy and Cirigliano, Vincenzo and Cohen, Saul and Gupta, Rajan and Joseph, Anosh and Lin, Huey-Wen and Yoon, Boram",
    collaboration = "PNDME",
    title = "{Iso-vector and Iso-scalar Tensor Charges of the Nucleon from Lattice QCD}",
    eprint = "1506.06411",
    archivePrefix = "arXiv",
    primaryClass = "hep-lat",
    reportNumber = "DESY-15-128, LA-UR-15-23801",
    doi = "10.1103/PhysRevD.92.094511",
    journal = "Phys. Rev. D",
    volume = "92",
    number = "9",
    pages = "094511",
    year = "2015"
}

@article{Gupta:2018lvp,
    author = "Gupta, Rajan and Yoon, Boram and Bhattacharya, Tanmoy and Cirigliano, Vincenzo and Jang, Yong-Chull and Lin, Huey-Wen",
    title = "{Flavor diagonal tensor charges of the nucleon from (2+1+1)-flavor lattice QCD}",
    eprint = "1808.07597",
    archivePrefix = "arXiv",
    primaryClass = "hep-lat",
    reportNumber = "LA-UR-18-28007",
    doi = "10.1103/PhysRevD.98.091501",
    journal = "Phys. Rev. D",
    volume = "98",
    number = "9",
    pages = "091501",
    year = "2018"
}

@article{Alexandrou:2019brg,
    author = "Alexandrou, C. and Bacchio, S. and Constantinou, M. and Finkenrath, J. and Hadjiyiannakou, K. and Jansen, K. and Koutsou, G. and Vaquero Aviles-Casco, A.",
    title = "{Nucleon axial, tensor, and scalar charges and $\sigma$-terms in lattice QCD}",
    eprint = "1909.00485",
    archivePrefix = "arXiv",
    primaryClass = "hep-lat",
    doi = "10.1103/PhysRevD.102.054517",
    journal = "Phys. Rev. D",
    volume = "102",
    number = "5",
    pages = "054517",
    year = "2020"
}

@article{Ledwig:2010tu,
    author = "Ledwig, Tim and Silva, Antonio and Kim, Hyun-Chul",
    title = "{Tensor charges and form factors of SU(3) baryons in the self-consistent SU(3) chiral quark-soliton model}",
    eprint = "1004.3612",
    archivePrefix = "arXiv",
    primaryClass = "hep-ph",
    reportNumber = "INHA-NTG-05-2010",
    doi = "10.1103/PhysRevD.82.034022",
    journal = "Phys. Rev. D",
    volume = "82",
    pages = "034022",
    year = "2010"
}

@article{Ledwig:2010zq,
    author = "Ledwig, Tim and Silva, Antonio and Kim, Hyun-Chul",
    title = "{Anomalous tensor magnetic moments and form factors of the proton in the self-consistent chiral quark-soliton model}",
    eprint = "1007.1355",
    archivePrefix = "arXiv",
    primaryClass = "hep-ph",
    reportNumber = "INHA-NTG-06-2010",
    doi = "10.1103/PhysRevD.82.054014",
    journal = "Phys. Rev. D",
    volume = "82",
    pages = "054014",
    year = "2010"
}

@article{Schmidt:1997vm,
    author = "Schmidt, Ivan and Soffer, Jacques",
    title = "{Melosh rotation and the nucleon tensor charge}",
    eprint = "hep-ph/9703411",
    archivePrefix = "arXiv",
    reportNumber = "CPT-97-P-3460",
    doi = "10.1016/S0370-2693(97)00737-5",
    journal = "Phys. Lett. B",
    volume = "407",
    pages = "331--334",
    year = "1997"
}

@article{Pasquini:2005dk,
    author = "Pasquini, B. and Pincetti, M. and Boffi, S.",
    title = "{Chiral-odd generalized parton distributions in constituent quark models}",
    eprint = "hep-ph/0510376",
    archivePrefix = "arXiv",
    doi = "10.1103/PhysRevD.72.094029",
    journal = "Phys. Rev. D",
    volume = "72",
    pages = "094029",
    year = "2005"
}

@article{Cloet:2007em,
    author = "Cloet, I. C. and Bentz, Wolfgang and Thomas, Anthony William",
    title = "{Transversity quark distributions in a covariant quark-diquark model}",
    eprint = "0708.3246",
    archivePrefix = "arXiv",
    primaryClass = "hep-ph",
    reportNumber = "PHY-11852-TH-2007, JLAB-THY-07-714",
    doi = "10.1016/j.physletb.2007.09.071",
    journal = "Phys. Lett. B",
    volume = "659",
    pages = "214--220",
    year = "2008"
}

@article{Maji:2017bcz,
    author = "Maji, Tanmay and Chakrabarti, Dipankar",
    title = "{Transverse structure of a proton in a light-front quark-diquark model}",
    eprint = "1702.04557",
    archivePrefix = "arXiv",
    primaryClass = "hep-ph",
    doi = "10.1103/PhysRevD.95.074009",
    journal = "Phys. Rev. D",
    volume = "95",
    number = "7",
    pages = "074009",
    year = "2017"
}

@article{Xu:2015kta,
    author = "Xu, Shu-Sheng and Chen, Chen and Cloet, Ian C. and Roberts, Craig D. and Segovia, Jorge and Zong, Hong-Shi",
    title = "{Contact-interaction Faddeev equation and, inter alia , proton tensor charges}",
    eprint = "1509.03311",
    archivePrefix = "arXiv",
    primaryClass = "nucl-th",
    doi = "10.1103/PhysRevD.92.114034",
    journal = "Phys. Rev. D",
    volume = "92",
    number = "11",
    pages = "114034",
    year = "2015"
}

@article{He:1994gz,
    author = "He, Han-xin and Ji, Xiang-Dong",
    title = "{The Nucleon's tensor charge}",
    eprint = "hep-ph/9412235",
    archivePrefix = "arXiv",
    reportNumber = "MIT-CTP-2380",
    doi = "10.1103/PhysRevD.52.2960",
    journal = "Phys. Rev. D",
    volume = "52",
    pages = "2960--2963",
    year = "1995"
}

@article{Aliev:2011ku,
    author = "Aliev, T. M. and Azizi, K. and Savci, M.",
    title = "{Nucleon tensor form factors induced by isovector and isoscalar currents in QCD}",
    eprint = "1108.2019",
    archivePrefix = "arXiv",
    primaryClass = "hep-ph",
    doi = "10.1103/PhysRevD.84.076005",
    journal = "Phys. Rev. D",
    volume = "84",
    pages = "076005",
    year = "2011"
}

@article{Diehl:2001pm,
    author = "Diehl, M.",
    title = "{Generalized parton distributions with helicity flip}",
    eprint = "hep-ph/0101335",
    archivePrefix = "arXiv",
    reportNumber = "DESY-01-009",
    doi = "10.1007/s100520100635",
    journal = "Eur. Phys. J. C",
    volume = "19",
    pages = "485--492",
    year = "2001"
}

@article{Diehl:2005jf,
    author = "Diehl, M. and Hagler, Ph.",
    title = "{Spin densities in the transverse plane and generalized transversity distributions}",
    eprint = "hep-ph/0504175",
    archivePrefix = "arXiv",
    reportNumber = "DESY-05-061",
    doi = "10.1140/epjc/s2005-02342-6",
    journal = "Eur. Phys. J. C",
    volume = "44",
    pages = "87--101",
    year = "2005"
}

@article{Burkardt:2005hp,
    author = "Burkardt, Matthias",
    title = "{Transverse deformation of parton distributions and transversity decomposition of angular momentum}",
    eprint = "hep-ph/0505189",
    archivePrefix = "arXiv",
    doi = "10.1103/PhysRevD.72.094020",
    journal = "Phys. Rev. D",
    volume = "72",
    pages = "094020",
    year = "2005"
}

@article{Gockeler:2006zu,
    author = {G{\"o}ckeler, M. and H{\"a}gler, Ph. and Horsley, R. and Nakamura, Y. and Pleiter, D. and Rakow, P. E. L. and Sch{\"a}fer, A. and Schierholz, G. and St{\"u}ben, H. and Zanotti, J. M.},
    collaboration = "QCDSF, UKQCD",
    title = "{Transverse spin structure of the nucleon from lattice QCD simulations}",
    eprint = "hep-lat/0612032",
    archivePrefix = "arXiv",
    reportNumber = "DESY-06-245, EDINBURGH-2006-41, TUM-T39-06-16",
    doi = "10.1103/PhysRevLett.98.222001",
    journal = "Phys. Rev. Lett.",
    volume = "98",
    pages = "222001",
    year = "2007"
}

@article{Ioffe:1981kw,
    author = "Ioffe, B. L.",
    title = "{Calculation of Baryon Masses in Quantum Chromodynamics}",
    reportNumber = "ITEP-10-1981, ITEP-92-1981",
    doi = "10.1016/0550-3213(81)90259-5",
    journal = "Nucl. Phys. B",
    volume = "188",
    pages = "317--341",
    year = "1981",
    note = "[Erratum: Nucl.Phys.B 191, 591--592 (1981)]"
}

@article{Braun:2006hz,
    author = "Braun, V. M. and Lenz, A. and Wittmann, M.",
    title = "{Nucleon Form Factors in QCD}",
    eprint = "hep-ph/0604050",
    archivePrefix = "arXiv",
    doi = "10.1103/PhysRevD.73.094019",
    journal = "Phys. Rev. D",
    volume = "73",
    pages = "094019",
    year = "2006"
}

@article{Diehl:1998kh,
    author = "Diehl, M. and Feldmann, T. and Jakob, R. and Kroll, P.",
    title = "{Linking parton distributions to form-factors and Compton scattering}",
    eprint = "hep-ph/9811253",
    archivePrefix = "arXiv",
    reportNumber = "DESY-98-172, WUB-98-37, FNT-T-98-10",
    doi = "10.1007/s100529901100",
    journal = "Eur. Phys. J. C",
    volume = "8",
    pages = "409--434",
    year = "1999"
}

@article{Braun:2000kw,
    author = "Braun, V. and Fries, R. J. and Mahnke, N. and Stein, E.",
    title = "{Higher twist distribution amplitudes of the nucleon in QCD}",
    eprint = "hep-ph/0007279",
    archivePrefix = "arXiv",
    reportNumber = "TPR-00-10",
    doi = "10.1016/S0550-3213(00)00516-2",
    journal = "Nucl. Phys. B",
    volume = "589",
    pages = "381--409",
    year = "2000",
    note = "[Erratum: Nucl.Phys.B 607, 433--433 (2001)]"
}

@article{ParticleDataGroup:2024cfk,
    author = "Navas, S. and others",
    collaboration = "Particle Data Group",
    title = "{Review of particle physics}",
    doi = "10.1103/PhysRevD.110.030001",
    journal = "Phys. Rev. D",
    volume = "110",
    number = "3",
    pages = "030001",
    year = "2024"
}

@article{Erkol:2011iw,
    author = "Erkol, Guray and Ozpineci, Altug",
    title = "{Tensor form factors of nucleon in QCD}",
    eprint = "1107.4584",
    archivePrefix = "arXiv",
    primaryClass = "hep-ph",
    doi = "10.1016/j.physletb.2011.09.089",
    journal = "Phys. Lett. B",
    volume = "704",
    pages = "551--558",
    year = "2011"
}

@article{Lenz:2009ar,
    author = "Lenz, Alexander and Gockeler, Meinulf and Kaltenbrunner, Thomas and Warkentin, Nikolaus",
    title = "{The Nucleon Distribution Amplitudes and their application to nucleon form factors and the $N \to \Delta$ transition at intermediate values of $Q^2$}",
    eprint = "0903.1723",
    archivePrefix = "arXiv",
    primaryClass = "hep-ph",
    doi = "10.1103/PhysRevD.79.093007",
    journal = "Phys. Rev. D",
    volume = "79",
    pages = "093007",
    year = "2009"
}

@article{Alexandrou:2022dtc,
    author = "Alexandrou, C. and others",
    title = "{Moments of the nucleon transverse quark spin densities using lattice QCD}",
    eprint = "2202.09871",
    archivePrefix = "arXiv",
    primaryClass = "hep-lat",
    doi = "10.1103/PhysRevD.107.054504",
    journal = "Phys. Rev. D",
    volume = "107",
    number = "5",
    pages = "054504",
    year = "2023"
}

@article{Ghim:2025gqo,
    author = "Ghim, Nam-Yong and Won, Ho-Yeon and Kim, June-Young and Kim, Hyun-Chul",
    title = "{Nucleon tensor form factors at large Nc}",
    eprint = "2501.12241",
    archivePrefix = "arXiv",
    primaryClass = "hep-ph",
    reportNumber = "INHA-NTG-02/2025, JLAB-THY-25-4080",
    doi = "10.1103/PhysRevD.111.074024",
    journal = "Phys. Rev. D",
    volume = "111",
    number = "7",
    pages = "074024",
    year = "2025"
}

@article{Ghim:2026qnp,
    author = "Ghim, Nam-Yong and Won, Ho-Yeon and Kim, June-Young and Kim, Hyun-Chul",
    title = "{Multipole structure of nucleon tensor form factors}",
    eprint = "2605.24435",
    archivePrefix = "arXiv",
    primaryClass = "hep-ph",
    reportNumber = "INHA-NTG-02/2026",
    doi = "10.1103/wtdr-92yh",
    journal = "Phys. Rev. D",
    volume = "114",
    number = "3",
    pages = "034016",
    year = "2026"
}

@article{Tezgin:2024tfh,
    author = "Tezgin, Kemal and Maynard, Brean and Schweitzer, Peter",
    title = "{Chiral-odd GPDs in the bag model}",
    eprint = "2404.11563",
    archivePrefix = "arXiv",
    primaryClass = "hep-ph",
    doi = "10.1103/PhysRevD.110.054028",
    journal = "Phys. Rev. D",
    volume = "110",
    number = "5",
    pages = "054028",
    year = "2024"
}

@article{Kim:2024ibz,
    author = "Kim, June-Young and Weiss, Christian",
    title = "{Chiral-odd generalized parton distributions in the large-Nc limit of QCD: Spin-flavor structure, polynomiality, and sum rules}",
    eprint = "2411.17634",
    archivePrefix = "arXiv",
    primaryClass = "hep-ph",
    reportNumber = "JLAB-THY-24-4239",
    doi = "10.1103/PhysRevD.111.074007",
    journal = "Phys. Rev. D",
    volume = "111",
    number = "7",
    pages = "074007",
    year = "2025"
}

@article{Kaur:2023lun,
    author = "Kaur, Satvir and Xu, Siqi and Mondal, Chandan and Zhao, Xingbo and Vary, James P.",
    collaboration = "BLFQ",
    title = "{Spatial imaging of proton via leading-twist nonskewed GPDs with basis light-front quantization}",
    eprint = "2307.09869",
    archivePrefix = "arXiv",
    primaryClass = "hep-ph",
    doi = "10.1103/PhysRevD.109.014015",
    journal = "Phys. Rev. D",
    volume = "109",
    number = "1",
    pages = "014015",
    year = "2024"
}

@article{Ozdem:2020vpt,
    author = {{\"O}zdem, Ula{\c{s}}},
    title = "{Isovector and isoscalar tensor form factors of $N(1535) \rightarrow N$ transition in light-cone QCD}",
    eprint = "2004.12312",
    archivePrefix = "arXiv",
    primaryClass = "hep-ph",
    doi = "10.1103/PhysRevD.102.014001",
    journal = "Phys. Rev. D",
    volume = "102",
    number = "1",
    pages = "014001",
    year = "2020"
}

@article{Ozdem:2021zbn,
    author = {{\"O}zdem, Ula{\c{s}}},
    title = "{Tensor form factors of N(1535) state via light-cone QCD}",
    doi = "10.1016/j.cjph.2021.04.018",
    journal = "Chin. J. Phys.",
    volume = "72",
    pages = "93--99",
    year = "2021"
}

@article{kucukarslan:2016xhx,
    author = "kucukarslan, A. and Ozdem, U. and Ozpineci, A.",
    title = "{Tensor form factors of the octet hyperons in QCD}",
    eprint = "1610.08358",
    archivePrefix = "arXiv",
    primaryClass = "hep-ph",
    doi = "10.1103/PhysRevD.94.094010",
    journal = "Phys. Rev. D",
    volume = "94",
    number = "9",
    pages = "094010",
    year = "2016"
}

@article{Asmaee:2025elo,
    author = "Asmaee, Z. and Hajirasouliha, N. and Azizi, K.",
    title = "{Tensor form factors of the {\ensuremath{\Delta}}+ baryon induced by isovector and isoscalar currents in QCD}",
    eprint = "2511.12360",
    archivePrefix = "arXiv",
    primaryClass = "hep-ph",
    doi = "10.1103/mlt1-gvnw",
    journal = "Phys. Rev. D",
    volume = "113",
    number = "5",
    pages = "054002",
    year = "2026"
}

@article{Asmaee:2026wrk,
    author = "Asmaee, Z. and Azizi, K.",
    title = "{Tensor form factors of decuplet hyperons in QCD}",
    eprint = "2601.17948",
    archivePrefix = "arXiv",
    primaryClass = "hep-ph",
    doi = "10.1103/zwcs-5qx3",
    journal = "Phys. Rev. D",
    volume = "113",
    number = "9",
    pages = "094020",
    year = "2026"
}

@article{Artru:1989zv,
    author = "Artru, Xavier and Mekhfi, Mustapha",
    title = "{Transversely Polarized Parton Densities, their Evolution and their Measurement}",
    reportNumber = "LPTHE-ORSAY-89-25",
    doi = "10.1007/BF01556280",
    journal = "Z. Phys. C",
    volume = "45",
    pages = "669",
    year = "1990"
}

@article{Cortes:1991ja,
    author = "Cortes, J. L. and Pire, B. and Ralston, J. P.",
    title = "{Measuring the transverse polarization of quarks in the proton}",
    reportNumber = "CPTH-A048-0491",
    doi = "10.1007/BF01565099",
    journal = "Z. Phys. C",
    volume = "55",
    pages = "409--416",
    year = "1992"
}

@article{Ivanov:2002jj,
    author = "Ivanov, D. Yu. and Pire, B. and Szymanowski, L. and Teryaev, O. V.",
    title = "{Probing chiral odd GPD's in diffractive electroproduction of two vector mesons}",
    eprint = "hep-ph/0209300",
    archivePrefix = "arXiv",
    reportNumber = "CPHT-RR-059-0602",
    doi = "10.1016/S0370-2693(02)02856-3",
    journal = "Phys. Lett. B",
    volume = "550",
    pages = "65--76",
    year = "2002"
}

@article{Boussarie:2016qop,
    author = "Boussarie, R. and Pire, B. and Szymanowski, L. and Wallon, S.",
    title = "{Exclusive photoproduction of a $\gamma\,\rho$ pair with a large invariant mass}",
    eprint = "1609.03830",
    archivePrefix = "arXiv",
    primaryClass = "hep-ph",
    reportNumber = "LPT-ORSAY-16-58, CPHT-RR038.072016, LPT-Orsay-16-58",
    doi = "10.1007/JHEP02(2017)054",
    journal = "JHEP",
    volume = "02",
    pages = "054",
    year = "2017",
    note = "[Erratum: JHEP 10, 029 (2018)]"
}

\end{document}